\documentclass[a4paper,11pt]{article}
\pdfoutput=1 % if your are submitting a pdflatex (i.e. if you have
\usepackage{jheppub} % for details on the use of the package, please
\usepackage[T1]{fontenc} % if needed

\usepackage{mathrsfs}

\newcommand\Ns{\mathcal{N}}
\newcommand\Js{\mathcal{J}}
\newcommand\Ls{\mathcal{L}}

\title{\boldmath A string-theoretical analog of non-maximal chaos in some Sachdev-Ye-Kitaev-like models}

\author[a]{Jin Chen,}
\author[b,c]{Chen Ma,}
\author[c,1]{and Chushun Tian\note{Corresponding author.}}
\affiliation[a]{Department of Physics, Xiamen University, Xiamen 361005, China}
\affiliation[b]{School of Physical Sciences, University of Chinese Academy of Sciences, Beijing 100049, China}
\affiliation[c]{CAS Key Laboratory of Theoretical Physics and Institute of Theoretical Physics, \\Chinese Academy of Sciences, Beijing 100190, China}
\emailAdd{zenofox@gmail.com}
\emailAdd{machen@itp.ac.cn}
\emailAdd{ct@itp.ac.cn}

\abstract{Very recently two of the present authors have studied the chaos exponent of some Sachdev-Ye-Kitaev (SYK)-like models for arbitrary interaction strength \cite{Ma22}. These models carry supersymmetric (SUSY) or SUSY-like structures. Namely, bosons and Majorana fermions are both present and each of them interacts with $(q-1)$ particles, but the model is not necessarily supersymmetric. It was found that the chaos exponents in different models, no matter whether they carry SUSY(-like) structures or not, all follow a universal single-parameter scaling law for large $q$, and by tuning that parameter continuously a flow from maximally chaotic to completely regular motion results. Here we report a string-theoretical analog of this chaotic phenomenon. Specifically, we consider closed string scattering near the two-sided AdS black hole, whose amplitude grows exponentially in the Schwarzschild time, with a rate determined by the Regge spin of the Pomeron exchanged during string scattering. We calculate the Pomeron Regge spin for strings of different types, including the bosonic string, the type II superstring and the heterotic superstring. We find that the Pomeron Regge spin also displays a single-parameter scaling behavior independent of string types, with the parameter depending on the string length and the length scale characterizing the spacetime curvature; moreover, the scaling function has the same limiting behaviors as that for the chaos exponent of SYK-like models. Remarkably, the flow from maximally chaotic to completely regular motion in SYK-like models corresponds to the flow of the Pomeron Regge spin from $2$ to $1$.}

\begin{document}
\maketitle
\flushbottom

\section{Introduction}
\label{sec:introduction}

\subsection{Motivations}
\label{sec:motivation}

Interests in complex quantum systems, that display chaoticity, are increasing rapidly. These systems can be divided into two classes. In Class I, that has long been at the heart of the studies of nonlinear dynamics, a quantum system has a well-defined classical limit \cite{Gutzwiller90,Izrailev90}, i.e. is the quantization of a classically chaotic system. Many chaotic phenomena occurring in such a system, ranging from spectral statistics \cite{Altland15} to quantum scars \cite{Bogolmony88,Berry89} and to weak localization \cite{Larkin96,Tian04,Tian05,Galitski17}, are tightly related to the instability of classical trajectory in phase space and thus of semiclassical origin. In Class II, a quantum system has no classical limit \cite{Berry77,Scharf89,Beenakker11,Tian14,Tian16,Tian20,Lai18,Kitaev15,Maldacena16,Maldacena17,Witten17,Gross21}. So chaotic phenomena occurring in such a system are of pure quantum origin, described in terms of continuity properties \cite{Guarneri10} and statistics \cite{Haake01} of spectra, the out-of-time-ordered correlator \cite{Kitaev15,Larkin69,Maldacena15}, and other characteristics. Representative systems in Class II carry various quantum degrees of freedom or structures, ranging from the $1/2$-spin \cite{Berry77,Scharf89,Beenakker11,Tian14,Tian16,Tian20,Lai18} to SUSY \cite{Maldacena17,Witten17} or string \cite{Gross21} structures.

Recent studies have discovered that simply endowing a dynamical system in Class I with an intrinsic quantum degree of freedom or structure --- thus brings that system to Class II --- can give rise to a wealth of unexpected quantum chaotic phenomena. For example, when a canonical model in chaotic dynamics, the so-called quantum kicked rotor \cite{Casati79,Izrailev90,Chirikov08,Fishman10} --- a rotating quantum particle subject to time-periodic ``kicks'', is endowed with a $1/2$-spin, its quantum dynamics undergoes dramatic changes \cite{Tian14,Tian16,Tian20}. Specifically, without that spin the rotor exhibits dynamical localization, which is an analog of Anderson localization in quantum disordered\footnote{Canonical studies of quantum chaos \cite{Gutzwiller90,Izrailev90} focus on an individual system --- in contrast to an ensemble of systems --- free of {\it extrinsic} randomness such as disorders. It remains a prominent unsolved problem in chaotic dynamics to determine whether and under what conditions a quantized classically chaotic system behaves as an ensemble of quantum disordered systems \cite{Casati84,Casati86,Jitomirskaya02,Tian04a}. Note that most analytic studies of the SYK model rely upon averaging over an ensemble of disorder realizations, that naturally invites the applications of diagrammatic and field-theoretical techniques \cite{Kitaev15,Maldacena16}, similar to the studies of Anderson localization in disordered systems \cite{Abrahams10}. But, as an individual system (e.g. a quantum kicked rotor, a quantum billiard, or a SYK model with specific disorder realization) is concerned, it is impossible to perform such disorder averaging.} systems \cite{Casati79,Izrailev90,Chirikov08,Fishman10,Fishman82,Raizen11,Garreau17}; when the spin is added, dynamical localization disappears and instead, tuning the effective Planck constant drives a sequence of dynamical localization-delocalization transitions, which was proved to be topological and be described by the same field theory as that describing the integer quantum Hall effect \cite{Tian14,Tian16}. It has also been shown \cite{Tian20}, with mathematical rigor, that adding a $1/2$-spin destroys dynamical
localization in the well-known Maryland model \cite{Fishman82a,Fishman84}, which finds its origin in the classical (Liouville) integrability \cite{Berry84}, and instead leads to a similar topological dynamical transition, with quantum chaotic diffusion appearing at the critical point.

Compared to the introduction of spin degrees of freedom, the introduction of SUSY(-like) structures is even more radical in quantum physics, especially in quantum chaos. (By SUSY-like it is meant that fermions and bosons are coupled but the system is not supersymmetric.) Nowadays SUSY is a building block of high-energy physics. One may expect that the adding of SUSY(-like) structures to quantum dynamical systems may also lead to unexpected quantum chaotic phenomena, similar to the adding of spin. Significant progresses achieved in this subject might potentially open a door to probe macroscopic effects of microscopic SUSY(-like) structures hidden in complex quantum systems, as chaos is the microscopic origin of many macroscopic phenomena. Such a route of searching SUSY effects indirectly is in sharp contrast to searching SUSY particles directly in colliders.

Systems that are particularly suitable for exploring this subject are SUSY(-like) variants of the SYK model \cite{Kitaev15,Sachdev93}. Some reasons are as follows. On the one hand, the SYK model is composed merely of interacting Majorana fermions and does not carry any SUSY(-like) structures. Its quantum chaotic properties have been well understood. In particular, in the strong interaction limit its chaos exponent satisfies the so-called maximal chaos bound, $2\pi/\beta$, with $\beta$ being the inverse temperature \cite{Maldacena15}. Remarkably, this maximal chaos can be interpreted in terms of Einstein gravitational scattering between two classical particles in the black hole background \cite{Kitaev15,Maldacena16}. This correspondence does not exist in canonical nonlinear dynamics, and builds up a deep connection between novel chaotic phenomena and black hole physics. On the other hand, the SYK model allows nice embedding of various SUSY(-like) structures, whose consequences are currently under investigations \cite{Ma22,Maldacena17,Witten17,Peng17,Li17,Yoon17,Liu18,Yoon18,Garcia18,Bulycheva18,Kato18,Ye20,He22,Mak21,Mak22,Volovich17,Ahn22,Zhao22,Peng21,Marcus19}. So it is natural to ask whether and how these structures influence system's chaotic behaviors.

In \cite{Ma22} two of us studied non-maximal chaos in some SUSY(-like) versions of the SYK model \cite{Maldacena17,Peng17,Marcus19}, where a particle (either boson or Majorana fermion) interacts with $(q-1)$ particles. It was found that, irrespective of the presence or the absence of SUSY(-like) structures and of the details of the latter structures (if present), the chaos exponents $\lambda_L$ of different models all follow a single-parameter scaling law, provided $q$ is large. Specifically, when $\lambda_L$ is rescaled by the maximal chaos bound, it obeys \footnote{We set Planck's constant to unity throughout this paper.}
\begin{equation}
\label{formula:lya_exponent}
\frac{\lambda_L}{2\pi/\beta}=v(\beta\Js).
%,
%\quad x = \pi v(x)/\cos\frac{\pi v(x)}{2},
\end{equation}
Here $v(x)$ is some scaling function (see Eq.~(\ref{eq:102}) below for its explicit form) which is universal, i.e. independent of the model as well as its parameters, and $\beta{\cal J}$ is the dimensionless scaling factor with ${\cal J}$ depending on the parameters of the Hamiltonian (e.g. $q$ and the coupling constant $J$). The constructions and the parameters of the model enter only into ${\cal J}$. Moreover, it was found that as the scaling factor $\beta{\cal J}$ increases continuously from zero to infinity, $\lambda_L$ increases monotonically from zero to the maximal chaos bound, and thus the motion becomes more and more chaotic and eventually the maximal chaos fully develops. By quantitative arguments it was further conjectured that the scaling law (\ref{formula:lya_exponent}) holds for more general $1$D SYK-like models with large $q$. These findings uncover some universal aspects of non-maximal chaos in SYK-like models.

\subsection{Statement of the problem and summary of the result}
\label{sec:problem}

At first glance, the finding of the scaling law seems discouraging, because the law does not forecast any chaotic phenomena that exist only in the presence of SUSY(-like) structures. However, it raises some interesting questions: Why is the chaos exponent immune to SUSY(-like) structures? Are there any deeper reasons for the scaling law and its independence of SUSY(-like) structures? In fact, although Eq.~(\ref{formula:lya_exponent}) gives $\lambda_L$ for arbitrary $\beta{\cal J}$, it was argued \cite{Ma22} that the scaling law and its independence of SUSY(-like) structures are rooted in the emergence of conformal symmetry in the limit $\beta{\cal J}\rightarrow\infty$. In other words, that symmetry, though existing only in the strong interaction limit, controls chaotic behaviors at {\it arbitrary} interaction strength, even when the strength is very small so that the motion is almost regular. So, since in that limit the chaos exponent, $2\pi/\beta$, has gravity correspondence, owing to the emergence of conformal symmetry \cite{Kitaev15,Maldacena15,Maldacena16}, one might naturally ask: {\it Whether and how can the scaling law and its universality be understood by gravity theories?} In this work we present an answer to this problem. For the SYK model with general $q$, Maldacena and Stanford proposed to interpret the leading correction to the maximal $\lambda_L$ as the stringy correction to the classical spin of graviton exchanged during bosonic string scattering in the black hole background \cite{Maldacena16}. Here we advance this idea substantially, and provide a string-theoretical analog of the scaling law (\ref{formula:lya_exponent}) and its independence of SUSY(-like) structures:
\begin{itemize}
  \item The scaling law (\ref{formula:lya_exponent}) corresponds to the following scaling law for the Regge spin $j_0$ of the Pomeron exchanged during closed string scattering near the two-sided AdS black hole:
    \begin{equation}\label{eq:60}
  j_0=1+v\left(\frac{4r^2(0)}{\alpha' \mu^2}\right).
\end{equation}
Here
%the scaling function ${\tilde v}(x)$ exhibits the same limiting behaviors as $v(x)$ for $x\rightarrow 0$ and $x\rightarrow \infty$, respectively.
$\alpha'$ gives the string tension and the string length, which are $1/(2\pi\alpha')$ and $\sqrt{\alpha^\prime}$, respectively. $r(0)\mu^{-1}$ is the characteristic length scale over which the spacetime is curved (see Eq.~(\ref{formula:scale_mu}) below for the definition). Thus strong (weak) chaos of SYK-like models correspond to the physics of short(long)-string scattering.
  \item That Eq.~(\ref{formula:lya_exponent}) is independent of SUSY(-like) structures finds its string-theoretical analog as that Eq.~(\ref{eq:60}) is independent of (super)string types.
\end{itemize}

We should emphasize that we do not aim at solving the difficult issue of the bulk-boundary duality, notably, the AdS-CFT or AdS-gauge duality, nor at establishing the gravity correspondence of the SYK model. Rather, we are discussing the possibility of understanding a non-maximally chaotic phenomenon in SYK-like models from the perspective of string scattering in the black hole background.

\subsection{Structure of the paper}
\label{sec:structure_paper}

In Sec.~\ref{sec:a_case_study} we review the scaling law (\ref{formula:lya_exponent}) and its universality. Section \ref{sec:general_description} is devoted to a general discussion on a possible gravity analog. In particular, we demonstrate the correspondence between the chaos exponent and the Regge spin of the Pomeron exchanged during string scattering in the black hole background, and the necessity of considering strings of different types. In Secs.~\ref{sec:pomeron_string_scattering}-\ref{sec:chaotic_exponent} we put the general discussion into practice. First of all, in Sec.~\ref{sec:pomeron_string_scattering} we use the technique developed by Brower, Polchinski, Strassler, and Tan (BPST) \cite{BPST07} to study the (tree-level) scattering amplitude for short strings propagating near a two-sided AdS black hole, and calculate the Pomeron exchanged during string scattering. We consider three different strings, the bosonic string, the type II superstring and the heterotic superstring. Then, in Sec.~\ref{sec:eikonal_phase_string_scattering} we use the results for the Pomeron to calculate the eikonal phase and the leading stringy correction to the Regge spin of Pomeron. In Sec.~\ref{sec:scattering_long_strings} we consider the long-string scattering in the same black hole background. The rigorous study of this subject is notoriously difficult. Thus we hypothesize a theoretical model. This allows us to calculate the Pomeron Regge spin in the long string limit.
%The result turns out to be different from that given by the super-Yang-Mills theory.
In Sec.~\ref{sec:chaotic_exponent} we use the results obtained in Secs.~\ref{sec:eikonal_phase_string_scattering} and \ref{sec:scattering_long_strings} to demonstrate how the scaling law (\ref{formula:lya_exponent}) and its universality can be interpreted from the perspective of string theory. We make concluding remarks in Sec.~\ref{sec:concluding_remarks}. Some technical details are shuffled to Appendixes \ref{sec:spin-j_laplacian} and \ref{sec:technical_details}.

\section{Review on the scaling law of non-maximal chaos exponent}
\label{sec:a_case_study}

In this section we review some results for the non-maximal chaos exponent and key steps of their derivations in four models: the SYK model, the ${\cal N}=1,\,2$ SUSY-SYK model and the so-called $(N|M)$-SYK model, where the interaction strength is arbitrary. $q$ is assumed to be large, and to be even for the first model and odd for the other three models. The scaling law (\ref{formula:lya_exponent}) and its universality are shown explicitly for these four models. A general mechanism for this non-maximally chaotic phenomenon in more general SYK-like models, that is rooted in the emergent conformal symmetry in the strong interaction limit, is discussed.

\begin{figure}[b]
\begin{center}
\includegraphics[width=8cm]{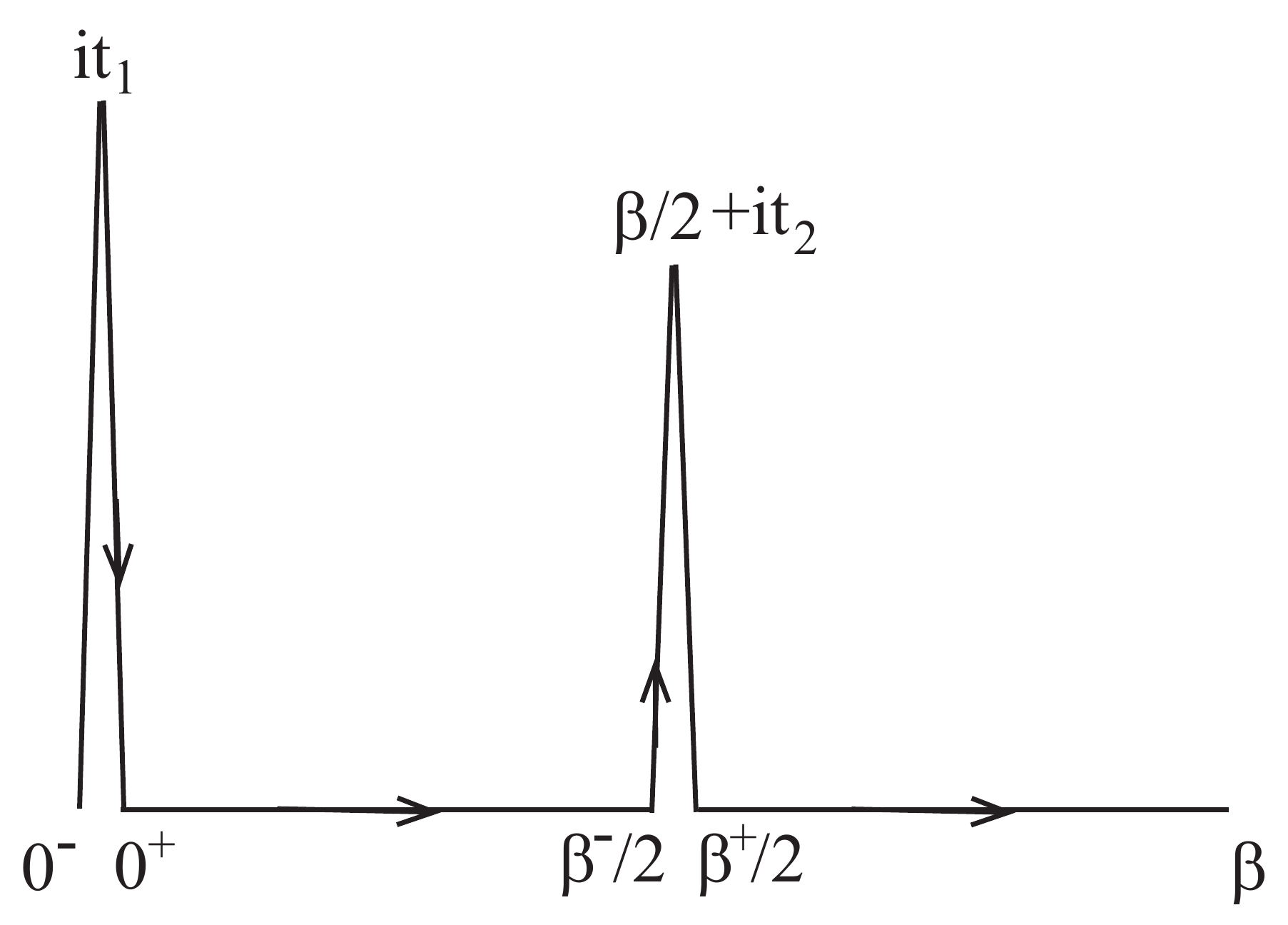}
\end{center}
\caption{The contour ${\cal C}$ in the definition of the double commutator, with the arrow indicating the contour direction.}
\label{fig:1}
\end{figure}

\subsection{SYK model}
\label{sec:non-maximal_chaos_SYK}

The SYK model \cite{Kitaev15,Sachdev93} consists of $N\gg 1$ Majorana fermions, each of which interacts with $(q-1)$ Majorana fermions. The coupling constants $C_{i_1\dots i_q}$ are antisymmetric with respect to $q$ fermion indices $i_1,\dots,i_q$. They are independent real Gaussian random variables, with zero mean and variance
\begin{equation}\label{eq:40}
 \overline{C_{i_1\dots i_q}^2} = \frac{(q-1)!J^2}{N^{q-1}},
\end{equation}
where the overline stands for the disorder average and $J$ for the interaction strength. Each Majorana fermion is described by a real fermionic field $\Psi_i(t)$, which depends only on the time $t$, and the Lagrangian of the model is
\begin{equation}\label{eq:62}
\Ls=\Psi_i \partial_t \Psi_i + \frac{i^{\frac{q}{2}}}{q!} C_{i_1\dots i_q} \Psi_{i_1}\dots\Psi_{i_q},
\end{equation}
where the second term accounts for the fermion interaction. Note that Einstein's summation convention is implied throughout this paper.

To probe chaotic phenomena we introduce the double commutator defined as
\begin{eqnarray}
\label{eq:63}
F(t_1;t_2)\equiv\frac{1}{N}\overline{\langle (\Psi_i(0^+) - \Psi_i(0^-)) \Psi_j(it_1)(\Psi_i({\beta^+\over 2}) - \Psi_i({\beta^-\over 2}))\Psi_j({\beta\over 2}+it_2)\rangle_\mathcal C},
\end{eqnarray}
where $t_{1,2}\in \mathbb{R}^+$ and $\langle\dots\rangle_{\mathcal C}$ denotes the functional average on the contour $\cal C$ in the complex time plane as shown in Fig.~\ref{fig:1}, i.e.
\begin{align}
\label{contour_average}
\langle \dots \rangle_{\mathcal C} \equiv \frac{\int D[\{\Psi_i\}] e^{\int_{\mathcal C} dt \mathcal L(\dots)}}{\int D[\{\Psi_i\}] e^{\int_{\mathcal C} dt \mathcal L}}.
\end{align}
This commutator was introduced in \cite{Witten17}, and is more parallel to the double commutator originally studied by Larkin and Ovchinnikov \cite{Larkin69}. It carries essentially the same information about chaotic behaviors as the out-of-time-ordered correlator (OTOC). Following \cite{Witten17} it can be shown that the double commutator satisfies the Bethe-Salpeter equation:
\begin{eqnarray}
\label{eq:64}
F(t_1;t_2)=F_0(t_1;t_2)+\int dt_3dt_4 K_{R}(t_1;t_2|t_3;t_4)F(t_3;t_4).
\end{eqnarray}
Here $F_0$ is the contribution of zero-rung diagram. The retarded kernel
\begin{equation}\label{eq:65}
  K_{R}(t_1;t_2|t_3;t_4)=\Theta(t_1-t_3)\Theta(t_2-t_4) W(t_3;t_4),
\end{equation}
where $\Theta$ is the Heaviside function and
\begin{align}
\label{eq:66}
W(t_3;t_4) = 2\left(\frac{\pi v}{\beta}\frac{1}{\cosh\frac{\pi v(t_3-t_4)}{\beta}}\right)^2,
\end{align}
with $v$ given by the equation:
\begin{align}
\label{eq:102}
x = \pi v/\cos \frac{\pi v}{2},\quad x=\beta\Js
\end{align}
and
\begin{equation}\label{eq:65}
  \Js=\frac{\sqrt{q}J}{2^{{q-1\over 2}}}.
\end{equation}
It is easy to see find the following limiting behaviors:
\begin{eqnarray}\label{eq:101}
  {v}(x)=\left\{\begin{array}{ll}
                        1-{2\over x}, & {\rm for}\, x\rightarrow\infty \\
                        {x\over \pi}, & {\rm for}\, x\rightarrow 0
                      \end{array}
  \right.
\end{eqnarray}
by using Eq.~(\ref{eq:102}).

Since chaotic phenomena occur at large $t_{1,2}$, we can ignore the first term on the right-hand side of Eq.~(\ref{eq:64}). By further performing the derivative: $\partial_{t_1}\partial_{t_2}$ on both sides, we obtain
\begin{align}
\label{formula:N=1_retarded_equation_diff_det}
\left(\partial_{t_1}\partial_{t_2} - {2\pi^2v^2\over\beta^2}{1\over \cosh^2{\pi v(t_1-t_2)\over \beta}}\right)F = 0.
\end{align}
Solving this equation, we find that \cite{Maldacena16}
\begin{align}
\label{eq:67}
F(t_1;t_2)=\frac{e^{{2\pi v\over \beta} \frac{t_1+t_2}{2}}}{\cosh \frac{\pi v(t_1-t_2)}{\beta}}.
\end{align}
In the special case of $t_1=t_2=t$, we have $
F(t;t) = e^{{2\pi v\over \beta}\,t}$. It grows exponentially in time with a rate of ${2\pi v\over \beta}$, which is a basic diagnosis of chaos \cite{Larkin69,Kitaev15}. Defining that rate as the chaos exponent, $\lambda_L$, immediately we find the scaling law (\ref{formula:lya_exponent}), and the scaling function $v(\beta{\Js})$ is given by Eq.~(\ref{eq:102}). Correspondingly, Eq.~(\ref{eq:67}) is rewritten as
\begin{align}
\label{eq:68}
F(t_1;t_2)=\frac{e^{\lambda_L\frac{t_1+t_2}{2}}}{\cosh \frac{\lambda_L(t_1-t_2)}{2}}.
\end{align}
That $\lambda_L<2\pi/\beta$ for finite $\Js$ signals the onset of non-maximal chaos.

Below we shall see that Eqs.~(\ref{eq:102}) and (\ref{formula:N=1_retarded_equation_diff_det}) hold for other SYK-like models, except the differences in the definition of $\Js$. This is a root of the scaling law as well as its independence of SUSY(-like) structures.

\subsection{$\Ns=1$ SUSY-SYK model}
\label{sec:N1_SUSY_SYK_model}

For the $\Ns=1$ SUSY-SYK model each Majorana fermion is represented by the superfield $\Psi_i(t,\theta)$ defined over the supersymmetric spacetime, which has a time direction, $t$, and a Grassmann direction, $\theta$. Its Lagrangian is \cite{Maldacena17}
\begin{equation}\label{formula:Lagrangian1}
\Ls=\int d\theta\left(\Psi_i D \Psi_i + \frac{i^{\frac{q-1}{2}}}{q!} C_{i_1\dots i_q} \Psi_{i_1}\dots\Psi_{i_q}\right),
\end{equation}
where similar to the SYK model the coupling constants $C_{i_1\dots i_q}$ are independent real Gaussian random variables, with zero mean and the variance given by Eq.~(\ref{eq:40}), and $D = \partial_{\theta} -i \theta \partial_{t}$ is the covariant derivative.

Non-maximal chaos of this model was studied in \cite{Ma22} with the help of the double commutator, defined as
\begin{eqnarray}
\label{doublecommutator}
F(\xi_1;\xi_2)\equiv\frac{1}{N}\overline{\langle (\Psi_i(0^+,0)\!-\!\Psi_i(0^-,0)) \Psi_j(it_1,\theta_1)(\Psi_i({\beta^+\over 2},0)\!-\! \Psi_i({\beta^-\over 2},0))\Psi_j({\beta\over 2} \!+\! it_2,\theta_2)\rangle_\mathcal C}\nonumber\\
\end{eqnarray}
with $\xi_i\equiv (t_i,\theta_i)$. It can be shown that the double commutator satisfies the following Bethe-Salpeter equation:
\begin{eqnarray}
\label{eq:35}
F(\xi_1;\xi_2)=F_0(\xi_1;\xi_2)+\int d\xi_4d\xi_3 K_{R}(\xi_1;\xi_2|
\xi_3;\xi_4)F(\xi_3;\xi_4),
\end{eqnarray}
with $d\xi_i\equiv dt_id\theta_i$. Here the order of the measures: $d\xi_4$ and $d\xi_3$ is in accordance with the path order. $F_0$ is the contribution of zero-rung diagram. The retarded kernel
\begin{equation}\label{eq:43}
  K_{R}(\xi_1;\xi_2|\xi_3;\xi_4)=-\Theta(t_1-t_3+i\theta_1\theta_3)\Theta(t_2-t_4+i\theta_2\theta_4) W(\xi_3;\xi_4),
\end{equation}
where
\begin{align}
\label{formula:N=1_potential_W}
W(\xi_3;\xi_4) = \frac{\pi v}{\beta}\frac{1}{\cosh\frac{\pi v(t_3-t_4)}{\beta} + \frac{\pi v}{\beta}\theta_3\theta_4},
\end{align}
with $v$ given by Eq.~(\ref{eq:102}) also but
\begin{align}
\label{formlua:N=1_large_q_v_constraint}
\Js=\frac{qJ^2}{2^{q-2}}
\end{align}
differing from $\Js$ in the SYK model.

For large $t_{1,2}$, again we can ignore the first term on the right-hand side of Eq.~(\ref{eq:35}). By further performing the derivative: $D_1D_2$ on both sides, where $D_1$ acts on $\xi_1$ and $D_2$ on $\xi_2$, we reduce Eq.~(\ref{eq:35}) to
\begin{eqnarray}
D_1 D_2F(\xi_1;\xi_2) = W(\xi_1;\xi_2)F(\xi_1;\xi_2).
\label{formula:N=1_retarded_equation_diff}
\end{eqnarray}
Its solution has the general form,
\begin{eqnarray}
\label{formula:N=1_vector_unknown}
      F(\xi_1;\xi_2)=F_\psi(t_1;t_2)+\theta_1\theta_2F_b(t_1;t_2).
\end{eqnarray}
With its substitution we find that Eq.~(\ref{formula:N=1_retarded_equation_diff}) reduces to Eq.~(\ref{formula:N=1_retarded_equation_diff_det}) again, i.e. $F_\psi$ satisfies Eq.~(\ref{formula:N=1_retarded_equation_diff_det}). As a result,
\begin{align}
\label{formula:N=1_solution}
F(\xi_1;\xi_2)=\frac{e^{\lambda_L \frac{t_1+t_2}{2}}}{\cosh \frac{\lambda_L(t_1-t_2)}{2}+\frac{\lambda_L}{2}\theta_1\theta_2},
\end{align}
and, again, $\lambda_L$ follows the scaling law (\ref{formula:lya_exponent}), with the scaling function $v(\beta{\Js})$ given by Eq.~(\ref{eq:102}). This result was reported in \cite{Ma22}. In the special case of $t_1=t_2=t$, this gives
\begin{equation}\label{eq:31}
  F(\xi_1;\xi_2)|_{t_1=t_2=t}=e^{\lambda_Lt}\left(1-{\lambda_L\over 2}\,\theta_1\theta_2\right).
\end{equation}
So $F$ is factorized: The first factor describes the exponential growth in $t$ and the seccond accounts for the SUSY effects. Therefore, introducing $\Ns=1$ SUSY has no impacts on the chaoticity.

\subsection{$\Ns=2$ SUSY-SYK model}
\label{sec:N2_SUSY_SYK_model}

For the $\Ns=2$ SUSY-SYK model the supersymmetric spacetime includes two Grassmann directions whose coordinates are denoted as $\theta,\bar\theta$. Moreover, a superfield is chiral or anti-chiral. In the former case, the superfield takes the general form: $\Psi(t,\theta,\bar\theta) = \Psi(t-i\theta\bar\theta,\theta)$ satisfying $\bar D\Psi=0$, while in the latter case the superfield, denoted as $\bar\Psi$, takes the general form: $\bar\Psi(t,\theta,\bar\theta)=\bar\Psi(t+i\theta\bar\theta,\bar\theta)$ satisfying $D\bar\Psi=0$. Here $D = \partial_\theta - i\bar\theta\partial_t$ and $\bar D = \partial_{\bar\theta} - i\theta\partial_t$. Such chirality enforces the following modifications of the Lagrangian,
\begin{eqnarray}
\label{formula:Lagrangian2}
  \Ls = \int d\bar{\theta}\bar{\Psi}_i D\Psi_i+\frac{i^{\frac{q-1}{2}}}{q!}\left(\int d{\theta}C_{i_1\dots i_q} \Psi_{i_1}\dots\Psi_{i_q}+\int d\bar{\theta}\bar{C}_{i_1\dots i_q} \bar\Psi_{i_1}\dots\bar\Psi_{i_q}\right),
\end{eqnarray}
where the second and third terms are (anti-)holomorphic superpotentials describing the interaction of $q$ chiral and anti-chiral Majorana fermions, respectively. Because the random coupling $C$'s are now complex numbers, Eq.~(\ref{eq:40}) is modified to be
\begin{equation}\label{eq:42}
 \overline{{\bar C}_{i_1\dots i_q}C_{i_1\dots i_q}} = \frac{(q-1)!J^2}{N^{q-1}},
\end{equation}
and all the other correlations of random couplings vanish.

Owing to the presence of the chirality, the arguments in the double commutator $F(\xi_1;\xi_2)$ correspond to distinct chirality. So, without loss of generality let $\xi_1=(s_1,\theta_1)$ and $\xi_2=({\bar s}_2,{\bar \theta}_2)$, where $s_1=t_1-i\theta_1{\bar \theta}_1$ and ${\bar s}_2=t_2+i\theta_2{\bar \theta}_2$, so that
\begin{equation}\label{eq:47}
  F(\xi_1;\xi_2)=F(s_1,\theta_1;{\bar s}_2,{\bar \theta}_2).
\end{equation}
It can be shown \cite{Ma22} that $F$ satisfies
\begin{eqnarray}
\label{eq:46}
F(s_1,\theta_1;{\bar s}_2,{\bar \theta}_2)&=&F_0(s_1,\theta_1;{\bar s}_2,{\bar \theta}_2)\nonumber\\
&&+\int ds_4d\theta_4d{\bar s}_3d{\bar \theta}_3 K_{R}(s_1,\theta_1;{\bar s}_2,{\bar \theta}_2|
{\bar s}_3,{\bar \theta}_3;s_4,\theta_4)F({\bar s}_3,{\bar \theta}_3;s_4,\theta_4).\quad
\end{eqnarray}
Here $F_0$ accounts for the zero-rung contribution. The order of the measures: $ds_4d\theta_4$ and $d{\bar s}_3d{\bar \theta}_3$ is again in accordance with the path order. The retarded kernel
\begin{eqnarray}\label{eq:51}
  &&K_{R}(s_1,\theta_1;{\bar s}_2,{\bar \theta}_2|
{\bar s}_3,{\bar \theta}_3;s_4,\theta_4)\nonumber\\
&=&-\frac{1}{4}\Theta\left(s_1-{\bar s}_3+2i\theta_1{\bar \theta}_3\right)\Theta\left({\bar s}_2-s_4+2i{\bar \theta}_2\theta_4\right) W({\bar s}_3,{\bar \theta}_3;s_4,\theta_4),
\end{eqnarray}
where
\begin{align}
\label{formula:N=2_potential_W}
W({\bar s}_3,{\bar \theta}_3;s_4,\theta_4) = \frac{2\pi v}{\beta}\frac{1}{\cosh\frac{\pi v({\bar s}_3-s_4)}{\beta} + \frac{2\pi v}{\beta}{\bar \theta}_3\theta_4},
\end{align}
with $v$ given by Eq.~(\ref{eq:102}) once more, but
\begin{align}
\label{eq:50}
\Js=\frac{2qJ^2}{4^{q-1}}
\end{align}
differing from $\Js$ in the SYK and the $\Ns=1$ SUSY-SYK model.

Similar to the derivations of Eq.~(\ref{formula:N=1_retarded_equation_diff}), we can obtain the differential equation:
\begin{align}
\label{formula:N=2_retarded_equation_diff}
D_1 {\bar D}_2 F(s_1,\theta_1;{\bar s}_2,{\bar \theta}_2) = W({\bar s}_1,{\bar \theta}_1;s_2,{\theta}_2) F({\bar s}_1,{\bar \theta}_1;s_2,{\theta}_2).
\end{align}
from Eq.~(\ref{eq:46}), where $D_1$ acts on the argument $(s_1,\theta_1)$ and ${\bar D}_2$ on $(s_2,{\bar \theta}_2)$. Importantly, comparing this equation with Eq.~(\ref{formula:N=1_retarded_equation_diff}), we see that here the arguments of $F$ on the left- and right-hand sides are different, inheriting the structures of Eq.~(\ref{eq:46}).

To solve Eq.~(\ref{formula:N=2_retarded_equation_diff}) we first perform the time translation:
\begin{eqnarray}
\label{eq:52}
  \left(
    \begin{array}{c}
      W(s_1,\theta_1;{\bar s}_2,{\bar \theta}_2) \\
      F(s_1,\theta_1;{\bar s}_2,{\bar \theta}_2) \\
    \end{array}
  \right)
   =e^{i(-\theta_1{\bar \theta}_1\partial_{t_1}+\theta_2{\bar \theta}_2\partial_{t_2})}\left(
    \begin{array}{c}
      W(t_1,\theta_1;t_2,{\bar \theta}_2) \\
      F(t_1,\theta_1;t_2,{\bar \theta}_2) \\
    \end{array}
  \right),
\end{eqnarray}
and likewise for $W({\bar s}_1,{\bar \theta}_1;s_2,{\theta}_2)$ and $F({\bar s}_1,{\bar \theta}_1;s_2,{\theta}_2)$. Then, $F(t_1,\theta_1;t_2,{\bar \theta}_2)$ takes the general form:
\begin{eqnarray}
\label{formula:N=2_vector_unknown}
      F(t_1,\theta_1;t_2,{\bar \theta}_2)   \equiv
      F_\psi(t_1;t_2) +\theta_1{\bar \theta}_2 F_b(t_1;t_2).
\end{eqnarray}
With its substitution we find that $F_\psi$ satisfies Eq.~(\ref{formula:N=1_retarded_equation_diff_det}) again. As a result,
\begin{align}
\label{formula:N=2_solution}
F(s_1,\theta_1;{\bar s}_2,{\bar \theta}_2)=\frac{e^{\lambda_L \frac{s_1+{\bar s}_2}{2}}}{\cosh \frac{\lambda_L (s_1-{\bar s}_2)}{2}+\lambda_L\theta_1{\bar \theta}_2},
\end{align}
and, again, $\lambda_L$ follows the scaling law (\ref{formula:lya_exponent}), with the scaling function $v(\beta{\Js})$ given by Eq.~(\ref{eq:102}). The scaling law for $\lambda_L$ in the $\Ns=2$ SUSY-SYK model was first found in \cite{Peng21}. In the special case of $t_1=t_2=t$, this gives
\begin{equation}\label{eq:57}
  F(\xi_1;\xi_2)|_{t_1=t_2=t}=e^{\lambda_Lt}\left(1-{\lambda_L\over 2} \left(i(\theta_1{\bar \theta}_1-\theta_2{\bar \theta}_2)+2\theta_1{\bar \theta}_2\right)\right)
\end{equation}
for large $t$. Like the $\Ns=1$ case $F$ is factorized also: $\Ns=2$ SUSY has no effects on the first factor that fully captures chaoticity, and modifies only the second factor.

\subsection{$(N|M)$-SYK model}
\label{sec:N_M_SYK_model}

In \cite{Marcus19} a variant of the SYK-model was introduced. It consists of $N$ Majorana fermions $\psi_i$ and $M$ bosons $b_\alpha$, with each boson interacting with $(q-1)$ Majorana fermions and each fermion interacting with a boson and $(q-2)$ fermions. Although this model carries some SUSY-like structures, it is not supersymmetric, even in the case of $N=M$. This model was dubbed the $(N|M)$-SYK model in \cite{Ma22}, with $N,M$ denoting the number of fermions and bosons, respectively, and the vertical bar denoting the SUSY-like structure of the model. The random coupling constants of this model take the form of $C_{\alpha [i_1\dots i_{q-1}]}$, where $\alpha$ labels the bosons and $i_1, \dots, i_{q-1}$ label the fermions. Unlike the SYK and the $\Ns=1,2$ SUSY-SYK model, here $C$ is antisymmetric only with respect to the fermionic indices $i_1, \dots, i_{q-1}$ in the square bracket of the subscript. The coupling constants are independent real Gaussian random variables, with zero mean and variance
\begin{align}
  \label{eq:69}
  \overline{C_{\alpha [i_1\dots i_{q-1}]}^2} = \frac{(q-1)! J^2 p}{N^{q-1}}, \quad p = \sqrt{\frac{N}{M}}.
\end{align}
The Lagrangian of this model reads
\begin{align}
  \label{eq:70}
  \mathcal L = -\psi_i \partial_\tau \psi_i + b_\alpha b_\alpha + \frac{i^\frac{q-1}{2}}{(q-1)!} C_{\alpha [i_1\dots i_{q-1}]} b_\alpha \psi_{i_1} \dots \psi_{i_{q-1}},
\end{align}
where the factor $(q-1)!$ arises from the antisymmetry with respect to $(q-1)$ fermionic indices. Note that a $\Ns=1$ SUSY-SYK model could be received by modifying Eq.~(\ref{eq:70}) with $p=1$, such that $C$ is antisymmetric in all of its $q$ indices.

For the $(N|M)$-SYK model, because of SUSY breaking we have to introduce two double commutators, one fermionic and the other bosonic. They are defined as \cite{Marcus19}
\begin{align}
  \label{eq:71}
    F_\psi(t_1;t_2) \equiv \frac{1}{N} \overline{\langle (\psi_i(0^+) - \psi_i(0^-)) \psi_j(it_1) (\psi_i({\beta^+\over 2})- \psi_i({\beta^-\over 2})) \psi_j(\frac{\beta}{2}+it_2)\rangle_{\mathcal C}}
\end{align}
and
\begin{align}
  \label{eq:72}
  F_b(t_1;t_2) \equiv \frac{p^2}{N} \overline{\langle (\psi_i(0^+) - \psi_i(0^-)) b_\alpha(it_1)(\psi_i({\beta^+\over 2})- \psi_i({\beta^-\over 2})) b_\alpha(\frac{\beta}{2}+it_2)\rangle_{\mathcal C}},
\end{align}
respectively. Non-maximal chaos of the $(N|M)$-SYK model was studied in \cite{Ma22} with the help of these two double commutators. In can be shown that they satisfy the following coupled Bethe-Salpeter equations:
\begin{align}
  \label{eq:73}
     F_\psi(t_1;t_2) = F_{\psi\,0}(t_1;t_2) + \int dt_3 dt_4 \Big(K_{11}(t_1;t_2|t_3;t_4)F_\psi(t_3;t_4) + K_{12}(t_1;t_2|t_3;t_4)F_b(t_3;t_4) \Big),
\end{align}
\begin{align}
  \label{eq:74}
 F_b(t_1;t_2) = F_{b\,0} (t_1;t_2) + \int dt_3 dt_4 K_{21}(t_1;t_2|t_3;t_4)F_\psi(t_3;t_4).
\end{align}
Here $F_{\psi\,0}$ and $F_{b\,0}$  account for the zero-rung contribution. $K_{11}$, $K_{12}$ and $K_{21}$ are the retarded kernels. They are found to be \cite{Ma22}
\begin{align}
\label{eq:77}
  K_{11}(t_1;t_2|t_3;t_4) &=-\frac{1}{p} \Theta(t_1-t_3)\Theta(t_2-t_4)W_b(t_3;t_4),\\
  \label{eq:75}
  K_{12}(t_1;t_2|t_3;t_4) &=\frac{1}{p} \Theta(t_1-t_3)\Theta(t_2-t_4)W_\psi(t_3;t_4),\\
  \label{eq:76}
  K_{21}(t_1;t_2|t_3;t_4) &=p\delta(t_1-t_3)\delta(t_2-t_4)W_\psi(t_3;t_4),
\end{align}
where
\begin{align}
  \label{eq:78}
  W_\psi(t_3;t_4) = \frac{\pi v}{\beta} \frac{1}{\cosh \frac{\pi v (t_3-t_4)}{\beta}}
\end{align}
and
\begin{align}
  \label{eq:79}
  W_b(t_3;t_4) = - \frac{p\pi^2 v^2 }{\beta^2} \frac{1}{\cosh^2 \frac{\pi v (t_3-t_4)}{\beta}},
\end{align}
with $v$ given by Eq.~(\ref{eq:102}) again, but
\begin{align}
\label{eq:80}
\Js=\frac{qJ^2}{2^{q-2}}
\end{align}
differing from $\Js$ in the SYK and the $\Ns=1,\,2$ SUSY-SYK model. From Eqs.~(\ref{eq:73}) and (\ref{eq:74}) we find that $F_\psi$ satisfies Eq.~(\ref{formula:N=1_retarded_equation_diff_det}) again. As a result,
\begin{align}
  \label{eq:81}
  F_\psi(t_1;t_2)=\frac{e^{\lambda_L\frac{t_1+t_2}{2}}}{\cosh \frac{\lambda_L(t_1-t_2)}{2}},\quad F_b(t_1;t_2) = \frac{p\lambda_L}{2}\frac{e^{\lambda_L \frac{t_1+t_2}{2}}}{\cosh^2 \frac{\lambda_L (t_1-t_2)}{2}}.
\end{align}
Here the chaos exponent $\lambda_L$ satisfies the same scaling law as that for the SYK and the $\Ns=1,\,2$ SUSY-SYK model. Interestingly, we can combine $F_{\psi,\, b}$ into a single function in the following way:
\begin{eqnarray}
  \label{eq:82}
  F_{(N|M)}(t_1,\theta_1;t_2,\theta_2)&\equiv&F_\psi(t_1;t_2)+\theta_1\theta_2F_b(t_1;t_2)\nonumber\\
  &=&\frac{e^{\lambda_L\frac{t_1+t_2}{2}}}{\cosh \frac{\lambda_L(t_1-t_2)}{2}-\frac{p\lambda_L}{2}\theta_1\theta_2}=F(t_1,i\sqrt{p}\theta_1;t_2,i\sqrt{p}\theta_2),
\end{eqnarray}
where $F$ in the second line is the double commutator for the $\Ns=1$ SUSY-SYK model, namely, Eq.~(\ref{formula:N=1_solution}). Most importantly, we have seen that the introduced SUSY-like structures have no effects on chaoticity, and enter only into the detailed expression of the double commutator.

\subsection{Possible roles of conformal symmetry in non-maximal chaos}
\label{sec:impact_non-maximal_chaos}

The results for the four models considered show that the non-maximal chaos displays remarkable universality, with the chaos exponent following the scaling law described by Eqs.~(\ref{formula:lya_exponent}) and (\ref{eq:102}). This raises many interesting questions, among which are the following: What are the physical reasons for this finding? To what extent is the scaling law universal? Some observations were made in \cite{Ma22} leading to potential answers. They suggest important roles played by the conformal symmetry emergent in the strong interaction limit, which is commonly conceived to be relevant only to the maximal chaos \cite{Kitaev15,Maldacena16,Stanford15}. Below we refine quantitative arguments made in \cite{Ma22}.

The starting point is that for different SYK-like models considered the fermionic double commutator, $F_\psi(t_1;t_2)$, fully captures chaotic behaviors. (We keep in mind that in the special case of original SYK model the double commutator is fermionic and has no bosonic counterpart.) So, it is sufficient to focus on $F_\psi$. In the strong interaction limit, by the emergent conformal symmetry and large $q$, we expect all $F_\psi$'s to satisfy the same equation:
\begin{align}
  \label{eq:83}
  \left(\partial_{t_1}\partial_{t_2} - \frac{2\pi^2}{\beta^2} \frac{1}{\cosh^2 \frac{\pi (t_1-t_2)}{\beta}}\right) F_\psi(t_1;t_2) = 0,\quad {\rm for}\,\, \beta{\cal J}\rightarrow\infty.
\end{align}
Its solution,
\begin{align}
  \label{eq:84}
  F_\psi(t_1;t_2) =\frac{e^{\frac{2\pi}{\beta} {t_1+t_2\over 2}}}{\cosh \frac{\pi(t_1-t_2)}{\beta}}\equiv {\tilde F}_\psi(t_1;t_2),
\end{align}
then describes the maximal chaoticity.

What happens away from the strong interaction limit? We make the following\\

\noindent{\it Hypothesis: Non-maximal chaos at arbitrary interaction can be obtained from maximal chaos in the strong interaction limit by proper scaling transformation: $t\rightarrow vt$.}\\

\noindent Applying this hypothesis, we obtain
\begin{align}
  \label{eq:85}
  \left(\partial_{t_1}\partial_{t_2} - \frac{2\pi^2 v^2}{\beta^2} \frac{1}{\cosh^2 \frac{\pi v(t_1-t_2)}{\beta}}\right) {\tilde F}_\psi(vt_1;vt_2) = 0.
\end{align}
From this the solution to Eq.~(\ref{formula:N=1_retarded_equation_diff_det}) is given by
\begin{align}
  \label{eq:86}
  F_\psi(t_1;t_2)={\tilde F}_\psi(vt_1;vt_2).
\end{align}
This relation suggests that the non-maximal chaos exponent at arbitrary $\beta\Js$ is the rescaling of the maximal chaos bound at $\beta\Js\rightarrow\infty$, i.e.
\begin{equation}\label{eq:87}
  \frac{2\pi}{\beta} \rightarrow \frac{2\pi v}{\beta} = \lambda_L.
\end{equation}

The next step is to determine the value of $v$. To this end we recall that in the strong interaction limit the emergence of conformal symmetry enforces the fermionic Euclidean propagator to take the general form:
\begin{align}
  \label{eq:88}
  G_\psi(\tau) = A\,{\rm sgn}(\tau)\left(\frac{\pi}{\beta \mathcal J \cos\left(\pi(\frac{1}{2} - \frac{|\tau|}{\beta})\right)}\right)^{\Delta_\psi}.
\end{align}
Here $A$ is a numerical factor involved in the definition of $G_\psi$ and is model-dependent. The power $\Delta_\psi$ and the expression of $\mathcal J$ in terms of $J$ and $q$ are model-dependent also. For arbitrary interaction strength, we apply the scaling transformation above, obtaining
\begin{align}
  \label{eq:89}
  G_\psi(\tau) \rightarrow A\,{\rm sgn}(\tau)\left(\frac{\pi v}{\beta \mathcal J \cos\left(\pi v(\frac{1}{2} - \frac{|\tau|}{\beta})\right)}\right)^{\Delta_\psi},
\end{align}
where $v$ in the numerator arises from the invariance under the scaling transformation. Because the value of $G_\psi$ at $\tau=0$ does not vary with the interaction strength in the large $q$ limit, we find immediately that $v$ is exactly determined by Eq.~(\ref{eq:102}). So we reproduce the scaling law (\ref{formula:lya_exponent}).

The analysis above suggests that non-maximal chaos at given interaction strength is determined by the conformal symmetry appearing only in the strong interaction limit, and the scaling transformation: $t\rightarrow vt$ that connects the chaos exponents at distinct interaction strength. As the analysis relies little on detailed modifications of the SYK model, except that a modified model must be $1$D and have large $q$. So we conjecture that such modified SYK models  would have a non-maximal chaos exponent obeying the scaling law (\ref{formula:lya_exponent}).

\section{Seeking for analogs in gravity theory}
\label{sec:general_description}

\subsection{Structureless point particles}
\label{sec:structureless_point_articles}

\begin{figure}[b]
\begin{center}
\includegraphics[width=11cm]{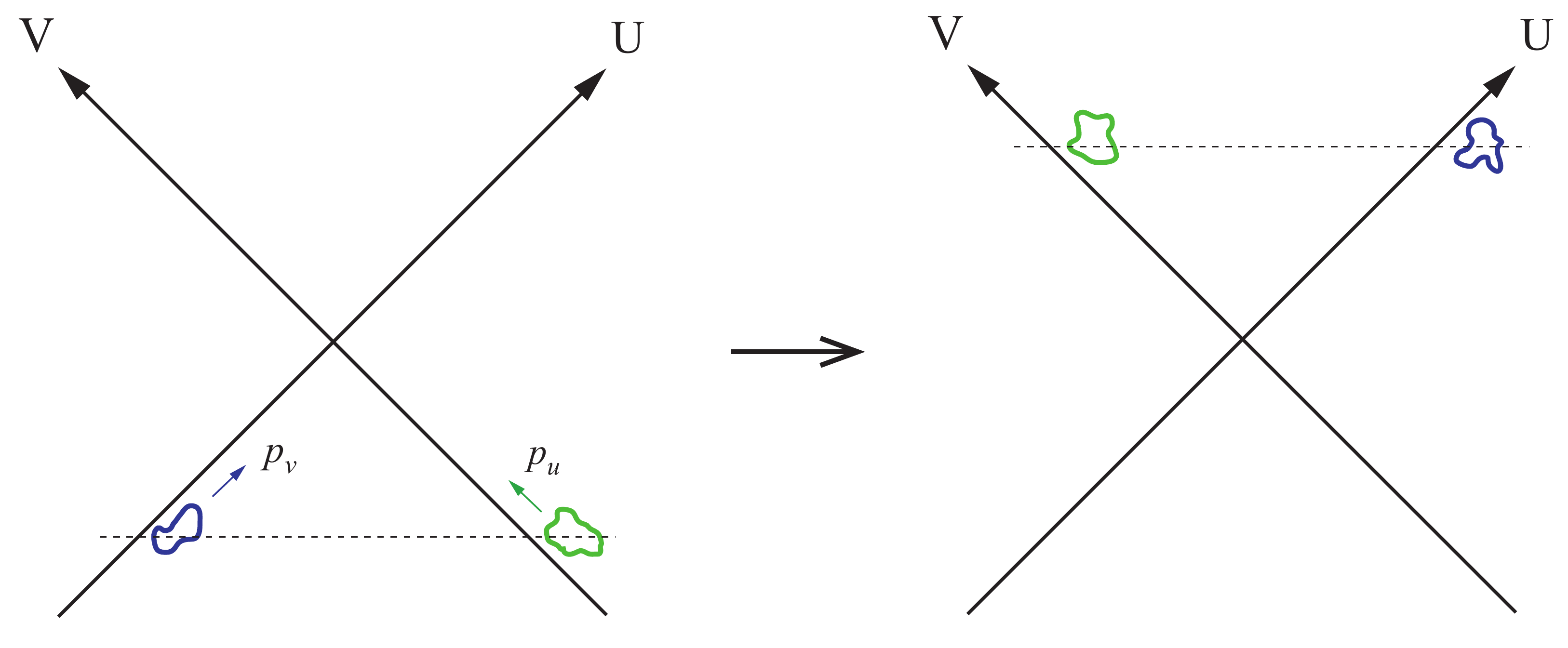}
\end{center}
\caption{High-energy scattering of two closed strings near a two-sided AdS black hole. The two strings in the incoming (left) or outgoing (right) state are in the same space-like slice represented by a dashed line. The maximal chaos bound arises when both strings shrink into point particles.}
\label{fig:2}
\end{figure}

We recall that for non-SUSY quantum mechanical systems, not limited to the SYK model, the maximal chaos bound can be interpreted in terms of Einstein gravitational scattering of two point particles \cite{Kitaev15,Maldacena15,Maldacena16,Stanford15}. The interpretation includes several steps (cf.~Fig.~\ref{fig:2}):
\begin{description}
  \item[(i)] The OTOC is expressed as the overlap of two states, $\Psi(t)$ and $\Psi'(t)$, on the boundary of a two-sided AdS black hole. The metric of this background spacetime, described in the Kruskal coordinates, is
\begin{align}
\label{formula:metric_black_hole}
ds^2 = -a(UV) dUdV + r^2(UV) \sum_{i=1}^{d-1} dX^i dX^i,
\end{align}
with $a,r>0$, which gives the metric coefficients:
\begin{equation}\label{eq:12}
  g_{UV}=g_{VU}=-\frac{a}{2},\quad g_{ii}=r^2, i=1,2,\cdots,d-1.
\end{equation}
The two horizons are located at $U=0$ and $V=0$, respectively, and intersect at $U=V=0$.
\item[(ii)] By letting two particles propagate from the boundary to a space-like slice in the bulk below the intersecting point, the state $\Psi(t)$ on the boundary is mapped onto a state in the bulk, which is a superposition of states (labelled by $I$) $\psi_{in,I}$, describing one particle traveling along the horizon of $U=0$ with a momentum $p_u$ and the other traveling along the horizon of $V=0$ with a momentum $p_v$. The two particles undergo Einstein gravitational scattering, with $\psi_{in,I}$ serving as the incoming state, and the outgoing state is denoted as $\psi_{out,J}$. Likewise, $\Psi'(t)$ has a bulk representation, which is the superposition of $\psi_{out,J}$.
\item[(iii)] So the OTOC is the superposition of scattering amplitudes $\langle\psi_{out,J}|\psi_{in,I}\rangle$. That is,
\begin{equation}\label{eq:58}
  {\rm OTOC}\,\sim\langle\Psi'(t)|\Psi(t)\rangle=\sum_{I,J} C_{IJ}\langle\psi_{out,J}|\psi_{in,I}\rangle,
\end{equation}
where $C_{IJ}$ are complex coefficients. Moreover, $\langle\psi_{out,J}|\psi_{in,I}\rangle$ is given by $e^{i\delta(s,b)}$, where the scattering phase $\delta$ depends on the Mandelstam-like square of the center-of-mass energy, $s\equiv 4p_u p_v/a(0)$, and the impact parameter $b$.
\item[(iv)] Near a black hole horizon, an outside Schwarzschild observer is accelerating, and the energy measured by him/her at a Schwarzschild time difference $t$ has a large relative boost $\sim e^{\frac{2\pi}{\beta}t}$, where $\beta$ is the inverse Hawking temperature. As such, $s$ grows exponentially in $t$ as
\begin{equation}\label{eq:34}
  s\sim \frac{1}{\beta^2} e^{\frac{2\pi}{\beta}t}.
\end{equation}
For such high-energy scattering the eikonal approximation works very well, and the eikonal phase displays the Regge behavior:
\begin{align}
\delta(s,b) \sim Gs^{j_0-1}.
\label{eq:32}
\end{align}
Here $G$ is Newton's constant and $j_0$ is the Regge spin, or more precisely the intercept of Regge trajectory, for which the general relativity gives the classical graviton spin $2$. Then we substitute $j_0=2$ into Eq.~(\ref{eq:32}), and expand the eikonal amplitude in $G$. With the substitution of the leading expansion into Eq.~(\ref{eq:58}), we obtain
\begin{eqnarray}\label{eq:59}
  {\rm OTOC}\,\stackrel{{\rm small}\, t}{\sim} Ge^{\frac{2\pi\left(j_0-1\right)}{\beta}t}\stackrel{j_0=2}{\longrightarrow}&Ge^{\frac{2\pi}{\beta}t},
\end{eqnarray}
which describes the exponential growth of OTOC at early time, with a rate $2\pi/\beta$. Thus Eq.~(\ref{eq:59}) serves as a gravity interpretation of the maximal chaos bound in non-SUSY quantum mechanical models.
\end{description}

\subsection{Point particles with SUSY(-like) structures and (super)string}
\label{sec:correspondence}

Now a question is, what happens if a particle acquires certain SUSY(-like) internal structures? If the particle remains point-like, we expect that the interpretation above remains valid and Eq.~(\ref{eq:59}) still holds. The reason is that Einstein theory for gravitational scattering is irrespective of point particle's internal structures. Consistent with this expectation, the SYK-like models studied in the last section do satisfy the maximal chaos bound in the strong interaction limit.

If the particle is no longer point-like, but instead is a string, Einstein theory ceases to work. In fact, a possible string realization of the SYK model has been proposed recently \cite{Verlinde21}. These motivate us to investigate the scattering of two strings in the same spacetime background. For scattering of two short bosonic closed strings this task was undertaken in \cite{Stanford15}. Specifically, when the two strings scatter they exchange a Pomeron, and the Pomeron Regge spin $j_0=2$ in Eq.~(\ref{eq:59}) was found to receive a negative stringy correction. This correction is argued to be the gravity interpretation of the leading correction to the maximal chaos exponent $2\pi/\beta$ for the SYK model but with general $q$ \cite{Maldacena16}.

In the remainder of this paper, we first show that this stringy correction is not changed, when the bosonic strings are replaced by type II and heterotic superstrings. This may be considered as a string-theoretical analog of the universality of the scaling law (\ref{formula:lya_exponent}) with respect to SUSY(-like) structures in the strong interaction regime (i.e. $\beta\Js\gg 1$). Next, we hypothesize a model for long (super)string scattering, and calculate the Pomeron Regge spin for long-string scattering. The results obtained for short- and long-string scattering give the single-parameter scaling law (\ref{eq:60}) for $j_0$ in the short- and long-string limit, respectively. They further suggest the chaos exponent-Pomeron Regge spin correspondence:
\begin{equation}\label{eq:19}
  {\lambda_L\over{2\pi/\beta}} \, \Leftrightarrow\, j_0-1
\end{equation}
for arbitrary interaction, with the parameter correspondence given by Eq.~(\ref{eq:27}) below. Moreover, the universality of the scaling law for the Pomeron Regge spin with respect to string types serves as an analog of the universality of the scaling law (\ref{formula:lya_exponent}) with respect to SUSY(-like) structures.

\section{(Super)string scattering amplitude and Pomeron}
\label{sec:pomeron_string_scattering}

In this section we will study the tree-level scattering amplitude of string scattering in the black hole background described by the metric (\ref{formula:metric_black_hole}), and derive the corresponding Pomeron operator by using the BPST technique. In this and the next section we shall consider closed strings and the strings are assumed to be short, for which the curvature effects of spacetime are weak. The quantitative requirement for the length of strings will be derived below and given in Eq.~(\ref{formula:range_small_string_mu}).

\subsection{Bosonic $(\Ns=\Ns^\prime=0)$ string}
\label{sec:bosonic_closed_string}

In this subsection, we shall derive the Pomeron operator for scattering of bosonic strings, for which the SUSY numbers $\Ns=\Ns^\prime=0$. For the following reasons we will present a self-contained and complete description of the theory for bosonic string scattering, although some derivations overlap with those in \cite{Stanford15}. First, it will make the paper more accessible to readers, especially to those who are unfamiliar with the BPST technique. Second, such a complete description paves the way to generalizing the derivations for bosonic strings to type II and heterotic strings, allowing subsequent investigations of whether non-maximal chaos is immune to the variations in SUSY numbers $(\Ns,\Ns^\prime)$. Third, we will derive the Pomeron completely within the framework of the worldsheet theory: This is a major difference from the previous study \cite{Stanford15} for bosonic string scattering in the black hole background; in that study some shock wave results \cite{tHooft1985} obtained from the general relativity were used, whereas in the present work those results are not needed.

Corresponding to the metric (\ref{formula:metric_black_hole}), the bosonic closed string, that have two sectors, namely, the left- and right-moving sector, is described by the worldsheet action:
\begin{equation}
\label{formula:bosonic_action_black_hole}
S=-\frac{1}{2\pi\alpha^\prime} \int dwd\bar w \left(r^2(UV) \sum_{i=1}^{d-1} \partial_w X^i \partial_{\bar w} X^i
- {a(UV)\over 2} (\partial_w U\partial_{\bar w} V + \partial_{\bar w}U\partial_w V)\right)
\end{equation}
in the conformal gauge. Here $(w,\bar w)$ are worldsheet coordinates, $U,V,\{X^i\}_{i=1}^{d-1}$ are the worldsheet fields that map the $2$D worldsheet onto the curved spacetime (\ref{formula:metric_black_hole}). Note that throughout this section $w$ can be anywhere on the entire complex plane $\mathbb{C}$. Provided that $r^2$ and $a$ are strictly constant, a flat spacetime results and Eq.~(\ref{formula:bosonic_action_black_hole}) describes a $2$D conformally invariant free field theory. Near the black hole the spacetime is curved over the scale of $r(0)\mu^{-1}$, with
\begin{align}
\label{formula:scale_mu}
\left.\mu^2 \equiv -\frac{d-1}{a}\partial_U\partial_V r^2\right|_{U=V=0}.
\end{align}
We are interested in the regime, where the string length is much smaller than this scale, i.e.
\begin{align}
\label{formula:range_small_string_mu}
\frac{\alpha^\prime \mu^2}{r^2(0)} \ll 1,
\end{align}
and in the leading curvature effects on the eikonal phase of Regge scattering. Therefore, we will study below the leading $\alpha^\prime \mu^2/r^2(0)$ corrections to the worldsheet conformal dimension of the operators involved in the operator product expansion (OPE), assuming that the operators and the OPE coefficients stay the same as those in the flat spacetime case \cite{Stanford15}.

In string theory it is well known that a physical string state is represented as a primary vertex operator. Specifically, for one string that propagates along the $V=0$ horizon, the vertex operators take the general form:
\begin{align}
\label{formula:vertex_operator_V}
\mathscr V_{2}(w,\bar w) &= g_{2}(U) T_{2}(X) e^{-i(p_v U + k_{2}\cdot X)},\notag\\
\mathscr V_{4}(w,\bar w) &= g_{4}(U) T_{4}(X) e^{+i(p_v U + k_{4}\cdot X)}
\end{align}
behind and in front of the $U=0$ horizon, respectively. Here all operators $U,X$ depend on $(w,\bar w)$. $T_{2,4}$ are the tensor part, whose explicit form depends on string states, e.g. tachyon, graviton, gauge boson, etc., and are only the function of transverse coordinates due to transverse polarization. $g_{2,4}$ are the envelop functions supported near $U=0$. $p_v$ is the momentum along the $V=0$ horizon and $k_{2,4}$ are the transverse momenta, with $p_{2,4}\gg k_{2,4}$. For another string that propagates along the $U=0$ horizon, similar to Eq.~(\ref{formula:vertex_operator_V}), we have the vertex operators
\begin{align}
\label{formula:vertex_operator_U}
\mathscr V_{1}(w,\bar w) &= g_{1}(V) T_{1}(X) e^{-i(p_u V + k_{1}\cdot X)},\notag\\
\mathscr V_{3}(w,\bar w) &= g_{3}(V) T_{3}(X) e^{+i(p_u V + k_{3}\cdot X)},
\end{align}
where all notations have the meanings similar to those in Eq.~(\ref{formula:vertex_operator_V}). Finally, $\mathscr V_{1,2,3,4}$ have conformal weight $(1,1)$. By the conformal symmetry we can bring three points in a four-point correlation function to $(0,0)$, $(1,1)$, and $(\infty,\infty)$ and fix them. Thus that correlation function depends only on one worldsheet coordinate $(w,\bar w)$. Taking this into account, we can write the tree-level scattering amplitude as
\begin{align}
\label{formula:amplitude_four_point_string}
\frac{g^2}{\pi} \int dwd\bar w \langle \mathscr V_4(0,0) \mathscr V_2(w,\bar w) \mathscr V_3(1,1) \mathscr V_1(\infty,\infty) \rangle,
\end{align}
where $\langle\dots\rangle$ stands for the worldsheet average corresponding to the action Eq.~(\ref{formula:bosonic_action_black_hole}), and $g$ is the gauge coupling \cite{Veneziano1987}.

To calculate Eq.~(\ref{formula:amplitude_four_point_string}) we perform OPE of $\mathscr V_4(0,0) \mathscr V_2(w,\bar w)$, obtaining
\begin{eqnarray}
\label{formula:OPE_pomeron}
\mathscr V_4(0,0)\mathscr V_2(w,\bar w)\stackrel{{\rm OPE}}{\sim}&c_{42} w^{L_0-2}\bar w^{{\tilde L}_0-2} g_2(U) g_4(U) e^{ik\cdot X - ip_v \cdot (\partial_w + \partial_{\bar w})U},\quad\,\,
\end{eqnarray}
where $L_0,{\tilde L}_0$ are the Virasoro left- and right-moving operator, respectively, $k=k_4-k_2$, the coefficient $c_{42}$ arises from contracting $T_2$ and $T_4$ which is the strength of coupling between the Pomeron and string states $2,4$, and the operators in the exponent are evaluated at $w = \bar w = 0$. In deriving Eq.~(\ref{formula:OPE_pomeron}) we have used the facts (i) that $\mathscr V_{2,4}$ have conformal dimension $(1,1)$ and (ii) that the product $g_2g_4$ is nonsingular at $w = \bar w = 0$ owing to the metric (\ref{formula:metric_black_hole}). Then, we diagonalize $L_0,{\tilde L}_0$ in the following basis with definite spin $j$:
\begin{align}
\label{formula:basis_spin_j}
\mathcal V (j) = (\partial_w U \partial_{\bar w} U)^{j\over 2} e^{ik\cdot X} g_2(U) g_4(U),
\end{align}
so that
\begin{align}
\label{formula:acting_L_0_on_the_basis}
L_0 \mathcal V(j) = {\tilde L}_0 \mathcal V(j) = (\partial_w U \partial_{\bar w}U)^{j\over 2} g_2(U) g_4(U) \left(\frac{j}{2} - \frac{\alpha^\prime \Delta_j}{4}\right) e^{ik\cdot X}.
\end{align}
Here $\Delta_j$ is the spin-$j$ generalization of the ordinary scalar Laplacian, $\Delta_0$, in curved spacetime \cite{BPST07}. The explicit form of $\Delta_{j>0}$, which will be derived in Appendix \ref{sec:spin-j_laplacian}, is
\begin{align}
\label{formula:Delta_j_sec}
\Delta_j = g_\perp^{j\over 4} \Delta_0 g_\perp^{-{j\over 4}} = g_\perp^{\frac{j-2}{4}} \Delta_2 g_\perp^{-\frac{j-2}{4}},
\end{align}
with
\begin{equation}
\label{formula:Delta_0_sec}
\Delta_0 = \frac{1}{r^2} \sum_{i=1}^{d-1} \partial_i^2- \frac{2}{a r^{d-1}}\left(\partial_U r^{d-1}\partial_V + \partial_V r^{d-1}\partial_U\right)
\end{equation}
and $g_\perp = r^{d-1}$. It turns out that all $\Delta_j$ reduce to the flat spacetime Laplacian in the flat spacetime limit. We may absorb the factors $g_\perp^{\pm j/4}$ in Eq.~(\ref{formula:Delta_j_sec}) into the factor $(\partial_w U \partial_{\bar w} U)^{j/2}$ in Eq.~(\ref{formula:acting_L_0_on_the_basis}), and rewrite Eq.~(\ref{formula:acting_L_0_on_the_basis}) as
\begin{eqnarray}
\label{formula:acting_L_0_on_basis_absorped}
L_0 \mathcal V(j)\!=\!{\tilde L}_0 \mathcal V(j)\!=\!\left(\left. g_\perp^{1\over 2}\right|_L \cdot \left. g_\perp^{-{1\over 2}}\right|_R \partial_w U \partial_{\bar w}U\right)^{j\over 2} g_2(U) g_4(U) \left(\frac{j}{2} - \frac{\alpha^\prime}{4} g_\perp^{-{1\over 2}} \Delta_2 g_\perp^{1\over 2}\right) e^{ik\cdot X},\,\,\quad
\end{eqnarray}
where $g_\perp^{1/2}|_{L(R)}$ is understood to act on the left (right) of $\Delta_2$.

Then we insert the OPE (\ref{formula:OPE_pomeron}) into Eq.~(\ref{formula:amplitude_four_point_string}) and carry out the integral over $w,\bar w$. For this purpose we note that in OPE only the terms that have matching powers of $w$ and $\bar w$ can contribute to the integral. Thus we can apply Eq.~(\ref{formula:acting_L_0_on_basis_absorped}) to Eq.~(\ref{formula:OPE_pomeron}), obtaining
\begin{eqnarray}
\label{OPE_Delta_2}
&&\mathscr V_4(0,0) \mathscr V_2(w,\bar w) \nonumber\\
&\stackrel{{\rm OPE}}{\sim}& c_{42} e^{-i \left.g_\perp^{1\over 2}\right|_L \cdot \left.g_\perp^{-{1\over 2}}\right|_R p_v (w\partial_w + \bar w \partial_{\bar w})U}
g_\perp^{-{1\over 2}} |w\bar w|^{-2-\frac{\alpha^\prime\Delta_2}{4}} g_\perp^{1\over 2}e^{ik\cdot X} g_2(U) g_4(U).
\end{eqnarray}
With the help of Eq.~(\ref{formula:Delta_0_sec}) we can compute the power $-\frac{\alpha^\prime\Delta_2}{4}$ explicitly. As a result,
\begin{eqnarray}
\label{OPE_mu_squared}
\mathscr V_4(0,0) \mathscr V_2(w,\bar w) \stackrel{{\rm OPE}}{\sim} c_{42} e^{-ip_v (w\partial_w + \bar w \partial_{\bar w})U}
|w\bar w|^{-2+\frac{\alpha^\prime(k^2+\mu^2)}{4r^2(0)}}e^{ik\cdot X} g_2(U) g_4(U).
\end{eqnarray}
Integrating out $w$,$\bar w$ gives the Pomeron operator
\begin{eqnarray}
\label{eq:18}
\int dwd\bar w \mathscr V_4(0,0) \mathscr V_2(w,\bar w)
= c_{42} \delta(U) \Pi_{(0, 0)}\left(\frac{\alpha^\prime t_\mu}{4}\right)e^{ik\cdot X} (p_v \partial_w U p_v\partial_{\bar w}U)^{1+\frac{\alpha^\prime t_\mu}{4}},
\end{eqnarray}
where
\begin{eqnarray}
\Pi_{(0, 0)}(x) = 2\pi \frac{\Gamma\left(-1 - x\right)}{\Gamma\left(2 + x\right)}e^{-i\pi \left(1 + x\right)}
\label{eq:14}
\end{eqnarray}
with $\Gamma$ being the Gamma function and the subscript $(0,0)$ denoting the SUSY numbers, $t_\mu = -(k^2+\mu^2)/r^2(0)$, and we have used the approximation \cite{Stanford15} $g_2(U) g_4(U) \simeq \delta(U)$.

\subsection{Type II ($\Ns=\Ns^\prime=1$) superstring}
\label{sec:typeII_superstring}

In this subsection we generalize the derivations of Pomeron operator for bosonic strings to type II superstrings, with the SUSY numbers $\Ns=\Ns^\prime=1$. This implies that one worldsheet fermion is added to the left-moving sector and the other to the right. So the worldsheet coordinate of the left-moving sector is $(w,\theta)$ and the right-moving $(\bar w, \bar\theta)$. Recall that $\theta,\bar\theta$ are Grassmann variables. With such supersymmetrization the operators: $U,V,X^i$ depend on $(w,\bar w,\theta,\bar\theta)$, and the derivatives $\partial_w, \partial_{\bar w}$ in Eq.~(\ref{formula:bosonic_action_black_hole}) are replaced by the covariant derivatives:
\begin{align}
\label{formula:typeII_SUSYderivatives}
  \partial_w \rightarrow D\equiv \partial_\theta + \theta\partial_w \qquad \partial_{\bar w} \rightarrow \bar D \equiv \partial_{\bar\theta} + \bar\theta \partial_{\bar w}.
\end{align}
With these replacements taken into account, we have the following supersymmetric worldsheet action:
\begin{align}
\label{formula:typeII_action_black_hole}
  S = -\frac{1}{2\pi\alpha^\prime} \int dwd\bar w d\theta d\bar\theta \left(r^2(UV)\sum_{i=1}^{d-1} DX^i \bar DX^i- {a(UV)\over 2}(D U\bar D V + D V \bar D U)\right)
\end{align}
as a descendant of the bosonic action (\ref{formula:bosonic_action_black_hole}). Were $r^2$ and $a$ strictly constant, Eq.~(\ref{formula:typeII_action_black_hole}) describes a $2$D superconformally invariant free field theory. The Pomeron operator for the type II superstring in the flat spacetime has been found in \cite{BPST07,Cheung10}. Below we derive the Pomeron operator for the type II superstring in the weakly curved spacetime described by the metric (\ref{formula:metric_black_hole}).

For superstrings, vertex operators in the Neveu-Schwarz sector, on which we focus throughout, can be presented in either the $-1$ or $0$ picture. In the $-1$ picture the worldsheet coordinates are not integrated out while in the $0$ picture are. Here we work in the $0$ picture, since the ensuing theory is parallel to the bosonic theory described in Sec.~\ref{sec:bosonic_closed_string}. In the present case, the operators (\ref{formula:vertex_operator_V}) and (\ref{formula:vertex_operator_U}) are replaced by
\begin{align}
\label{formula:vertex_operator_SUSY_general}
  \mathscr V_j (w,\bar w) = \int d\theta d\bar\theta \tilde{\mathscr V}_j(w,\bar w,\theta,\bar\theta), \quad j=1,2,3,4
\end{align}
where $\tilde{\mathscr V}_j$ has exactly the same form as those in Eqs.~(\ref{formula:vertex_operator_V}) and (\ref{formula:vertex_operator_U}), except that $U,V,X^i$ now depend on $(w,\bar w,\theta,\bar\theta)$, and has conformal dimension $(\frac{1}{2},\frac{1}{2})$. Moreover, the left- and right-moving sector of $\tilde{\mathscr V}_j$ must have the correct Glizzi-Scherk-Olive (GSO) projection \cite{Glizzi77}, so that each sector of $\tilde{\mathscr V}_j$ represents a fermion and consequently each sector of $\mathscr V_j$ represents a boson. It is worth mentioning that this projection is not essential to the expression of Pomeron operator \cite{BPST07}.

Next, we wish to perform the OPE of $\tilde{\mathscr V}_4(0,0,\theta^\prime,\bar\theta^\prime)\tilde{\mathscr V}_2(w,\bar w,\theta,\bar\theta)$. Again the tensor part of $\tilde{\mathscr V}_4,\tilde{\mathscr V}_2$ leads only to an unimportant overall coupling constant. Then, similar to Eq.~(\ref{formula:OPE_pomeron}) this operator product can be expanded in terms of the following operator:
\begin{align}
\label{formula:typeII_plane_wave}
  e^{ik\cdot X - ip_v \left((w\partial_w + \bar w \partial_{\bar w}) + (\theta-\theta^\prime)\partial_\theta + (\bar\theta - \bar\theta^\prime)\partial_{\bar\theta^\prime}\right) U},
\end{align}
where $X^i$ and the derivatives of $U$ are evaluated at the worldsheet coordinate $(0,0,\theta^\prime,\bar\theta^\prime)$. Upon performing the shift: $w\rightarrow w-\theta\theta^\prime, \bar w \rightarrow \bar w - \bar\theta\bar\theta^\prime$, we rewrite Eq.~(\ref{formula:typeII_plane_wave}) as
\begin{align}
\label{formula:typeII_plane_wave_shifted}
  e^{ik\cdot X - ip_v \left((w-\theta\theta^\prime)\partial_w + (\bar w - \bar\theta\bar\theta^\prime)\partial_{\bar w} + (\theta-\theta^\prime) D + (\bar\theta-\bar\theta^\prime){\bar D}\right)U}.
\end{align}
In these bases the OPE of $\tilde{\mathscr V}_4(0,0,\theta^\prime,\bar\theta^\prime)\tilde{\mathscr V}_2(w,\bar w,\theta,\bar\theta)$ takes the following form:
\begin{eqnarray}
\label{formula:typeII_OPE}
  \tilde{\mathscr V}_4(0,0,\theta^\prime,\bar\theta^\prime)\tilde{\mathscr V}_2(w,\bar w,\theta,\bar\theta)
  &\stackrel{{\rm OPE}}{\sim}& c_{42}(w-\theta\theta^\prime)^{L_0-1} (\bar w - \bar\theta\bar\theta^\prime)^{\tilde L_0 - 1}g_2(U) g_4(U) \nonumber\\
  &\times& e^{ik\cdot X - ip_v \left(
  %(w-\theta\theta^\prime)
  \partial_w +
  %(\bar w - \bar\theta\bar\theta^\prime)
  \partial_{\bar w} + (\theta-\theta^\prime) D + (\bar\theta-\bar\theta^\prime){\bar D}\right)U}.
\end{eqnarray}
When we perform the expansion in $\partial_w U, \partial_{\bar w}U$, only terms of the form: $(\partial_w U \partial_{\bar w}U)^{j/2}$ survive. Therefore, Eqs.~(\ref{formula:basis_spin_j}), (\ref{formula:acting_L_0_on_the_basis}) and (\ref{formula:acting_L_0_on_basis_absorped}) still apply. With their substitution Eq.~(\ref{formula:typeII_OPE}) reduces to
\begin{eqnarray}
\label{formula:typeII_formula:OPE_mu_squared}
  \tilde{\mathscr V}_4(0,0,\theta^\prime,\bar\theta^\prime)\tilde{\mathscr V}_2(w,\bar w,\theta,\bar\theta)
  &\stackrel{{\rm OPE}}{\sim}& c_{42}e^{ik\cdot X-ip_v \left((w-\theta\theta^\prime)\partial_w + (\bar w - \bar\theta\bar\theta^\prime)\partial_{\bar w} + (\theta-\theta^\prime) D + (\bar\theta-\bar\theta^\prime){\bar D}\right)U}\nonumber\\
  &\times &|(w-\theta\theta^\prime)(\bar w - \bar\theta\bar\theta^\prime)|^{-1+\frac{\alpha^\prime(k^2+\mu^2)}{4r^2(0)}} g_2(U) g_4(U).
\end{eqnarray}
Then we carry out the integral over Grassmann variables $\theta,\theta^\prime,\bar\theta,\bar\theta^\prime$. Because we are interested in the terms which are most singular at $w,\bar w \simeq 0$, we need to pick up only the terms resulting from the ($\theta,\theta^\prime,\bar\theta,\bar\theta^\prime$) expansion of the OPE coefficients, i.e.
\begin{align}
\label{formula:typeII_OPE_Grassmann_integrated}
  &\int d\theta d\bar\theta d\theta^\prime d\bar\theta^\prime~\tilde{\mathscr V}_4(0,0,\theta^\prime,\bar\theta^\prime)\tilde{\mathscr V}_2(w,\bar w,\theta,\bar\theta)\notag\\
  \sim & c_{42}\left(1 - \frac{\alpha^\prime(k^2+\mu^2)}{4r^2(0)}\right)^2 |w\bar w|^{-2 + \frac{\alpha^\prime(k^2+\mu^2)}{4r^2(0)}} e^{ik\cdot X-ip_v(w\partial_w + \bar w \partial_{\bar w})U} g_2(U)g_4(U),
\end{align}
where all operators are evaluated at $(0,0,0,0)$. Further integrating out $w,\bar w$ gives the Pomeron operator
\begin{align}
\label{formula:typeII_Pomeron_operator}
  \int dw d\bar w~\mathscr V_4(0,0) \mathscr V_2(w,\bar w)
  =c_{42}\delta(U)\Pi_{(1,1)}\left(\frac{\alpha^\prime t_\mu}{4}\right) e^{ik\cdot X} \left(p_v \partial_w U p_v \partial_{\bar w}U\right)^{1+\frac{\alpha^\prime t_\mu}{4}},
\end{align}
where
\begin{align}
\label{formula:typeII_Pi_function}
  \Pi_{(1,1)}\left(\frac{\alpha^\prime t_\mu}{4}\right) = \left(1 + \frac{\alpha^\prime t_\mu}{4}\right)^2\Pi_{(0,0)}\left(\frac{\alpha^\prime t_\mu}{4}\right),
\end{align}
with the factor in front of $\Pi_{(0,0)}$ inheriting from that in Eq.~(\ref{formula:typeII_OPE_Grassmann_integrated}) arising from the expansion in Grassmann variables, and the subscript $(1,1)$ denoting the SUSY numbers.

\subsection{Heterotic ($\Ns=1, \Ns^\prime=0$) superstring}
\label{sec:heterotic_superstring}

In this subsection we generalize the derivations of Pomeron operator to a heterotic string, with $\Ns=1,\,\Ns^\prime=0$. This implies that we add only a worldsheet fermion to the left-moving sector. So the worldsheet coordinate of this sector is modified to $(w,\theta)$, while that of the other sector stays the same which is $\bar w$. Correspondingly, the operators depend on $(w,\bar w,\theta)$ and the derivative $\partial_w$ is replaced by $D$ defined in Eq.~(\ref{formula:typeII_SUSYderivatives}) while $\partial_{\bar w}$ is not modified. The heterotic string is described by the worldsheet action:
\begin{align}
\label{formula:heterotic_action_black_hole}
  S= -\frac{1}{2\pi\alpha^\prime} \int dwd\bar w d\theta\left(r^2(UV) \sum_{i=1}^{d-1} DX^i\partial_{\bar w}X^i- {a(UV)\over 2}(DU\partial_{\bar w}V + DV\partial_{\bar w}U)\right),
\end{align}
which may be considered as the hybridization of the bosonic string action (\ref{formula:bosonic_action_black_hole}) and the type II superstring action. This is an example of the general $\sigma$ model action for the heterotic string without right-moving fermionic superfields \cite{Witten85}. In the flat spacetime, for which $r^2$ and $a$ are strictly constant, the Pomeron operator was found in \cite{Cheung10}. Below we derive the Pomeron operator for the heterotic string in the weakly curved spacetime described by the metric (\ref{formula:metric_black_hole}).

Because the left-moving sector is supersymmetric, the vertex operator associated with this sector must be GSO-allowed. Working in the $0$ picture, the operators in Eqs.~(\ref{formula:vertex_operator_V}) and (\ref{formula:vertex_operator_U}) are modified to
\begin{align}
\label{formula:heterotic_vertex_operator}
  \mathscr V_j(w,\bar w) = \int d\theta \tilde{\mathscr V}_j(w,\bar w,\theta), \quad j=1,2,3,4
\end{align}
where $\tilde{\mathscr V}_j$ has exactly the same form as those in Eqs.~(\ref{formula:vertex_operator_V}) and (\ref{formula:vertex_operator_U}), except that $U,V,X^i$ now depend on $(w,\bar w,\theta)$, and has conformal dimension $(\frac{1}{2},1)$.

Next, we wish to perform the OPE of $\tilde{\mathscr V}_4(0,0,\theta^\prime) \tilde{\mathscr V}_2(w,\bar w,\theta)$. Similar to Eqs.~(\ref{formula:OPE_pomeron}) and (\ref{formula:typeII_OPE}), the tensor part of $\tilde{\mathscr V}_4$, $\tilde{\mathscr V}_2$ leads only to an unimportant overall coupling constant, this operator product can be expanded in terms of the following operator:
\begin{align}
\label{formula:heterotic_plane_wave}
  e^{ik\cdot X - ip_v \left((w\partial_w + \bar w \partial_{\bar w}) + (\theta-\theta^\prime)\partial_{\theta^\prime}\right) U},
\end{align}
where $X^i$ and the derivatives of $U$ are evaluated at $(0,0,\theta^\prime)$. Upon performing the shift: $w\rightarrow w-\theta\theta^\prime$, we rewrite Eq.~(\ref{formula:heterotic_plane_wave}) as
\begin{align}
\label{formula:heterotic_plane_wave_shifted}
  e^{ik\cdot X - ip_v \left((w-\theta\theta^\prime) \partial_w + \bar w \partial_{\bar w}) + (\theta-\theta^\prime)D \right)U}.
\end{align}
In these bases the OPE of $\tilde{\mathscr V}_4(0,0,\theta^\prime) \tilde{\mathscr V}_2(w,\bar w,\theta)$ takes the following form:
\begin{align}
\label{formula:heterotic_OPE}
  &\tilde{\mathscr V}_4(0,0,\theta^\prime) \tilde{\mathscr V}_2(w,\bar w,\theta)\notag\\
  \stackrel{{\rm OPE}}{\sim}& c_{42}e^{ik\cdot X - ip_v \left(
  %(w-\theta\theta^\prime)
  \partial_w +
  %\bar w
  \partial_{\bar w} + (\theta-\theta^\prime) D\right) U} (w-\theta\theta^\prime)^{L_0-1}\bar w^{{\tilde L}_0-2}g_2(U)g_4(U).
\end{align}
Like in the cases of $\Ns=\Ns^\prime=0,1$, when we perform the expansion in $\partial_w U$, $\partial_{\bar w} U$, we can still use Eqs.~(\ref{formula:basis_spin_j}), (\ref{formula:acting_L_0_on_the_basis}), and (\ref{formula:acting_L_0_on_basis_absorped}) to reduce Eq.~(\ref{formula:heterotic_OPE}) to
\begin{align}
\label{formula:heterotic_formula:OPE_mu_squared}
  \tilde{\mathscr V}_4(0,0,\theta^\prime) \tilde{\mathscr V}_2(w,\bar w,\theta)
  \stackrel{{\rm OPE}}{\sim}& c_{42} e^{ik\cdot X-ip_v \left((w-\theta\theta^\prime)\partial_w + \bar w \partial_{\bar w} + (\theta-\theta^\prime) D\right) U}\notag\\
  \times& (w-\theta\theta^\prime)^{-1+\frac{\alpha^\prime(k^2+\mu^2)}{4r^2(0)}} \bar w^{-2+\frac{\alpha^\prime(k^2+\mu^2)}{4r^2(0)}}g_2(U)g_4(U).
\end{align}
Then we carry out the integral over $\theta,\theta^\prime$. Because we are interested in the terms which are most singular at $w,\bar w \simeq 0$, we need to pick up only the terms resulting from the ($\theta,\theta^\prime$) expansion of the OPE coefficients, i.e.
\begin{align}
\label{formula:heterotic_OPE_Grassmann_integrated}
  &\int d\theta d\theta^\prime \tilde{\mathscr V}_4(0,0,\theta^\prime) \tilde{\mathscr V}_2(w,\bar w,\theta)\notag\\
  \sim&c_{42}\left(1-\frac{\alpha^\prime(k^2+\mu^2)}{4r^2(0)}\right)|w\bar w|^{-2 + \frac{\alpha^\prime(k^2+\mu^2)}{4r^2(0)}} e^{ik\cdot X-ip_v (w\partial_w + \bar w \partial_{\bar w}) U}g_2(U)g_4(U),
\end{align}
where all operators are evaluated at $(0,0,0)$. Further integrating out $w,\bar w$ gives the Pomeron operator
\begin{align}
\label{formula:heterotic_Pomeron_operator}
  \int dw d\bar w \mathscr V_4(0,0) \mathscr V_2(w,\bar w)=c_{42} \delta(U) \Pi_{(1,0)}\left(\frac{\alpha^\prime t_\mu}{4}\right) e^{ik\cdot X} \left(p_v\partial_w U p_v \partial_{\bar w} U\right)^{1+\frac{\alpha^\prime t_\mu}{4}},
\end{align}
where
\begin{align}
\label{formula:heterotic_Pi_function}
  \Pi_{(1,0)}\left(\frac{\alpha^\prime t_\mu}{4}\right) = \left(1 + \frac{\alpha^\prime t_\mu}{4}\right) \Pi_{(0,0)}\left(\frac{\alpha^\prime t_\mu}{4}\right),
\end{align}
with the factor in front of $\Pi_{(0,0)}$ inheriting from that in Eq.~(\ref{formula:heterotic_OPE_Grassmann_integrated}) arising from the expansion in Grassmann variables, and the subscript $(1,0)$ denoting the SUSY numbers.

\subsection{General structures of Pomeron operator}
\label{sec:general_expression}

We may combine the results obtained in Secs.~\ref{sec:bosonic_closed_string}-\ref{sec:heterotic_superstring} together into a single expression, which allows us to investigate the general structures of Pomeron at different $(\Ns,\Ns')$. To this end, we rewrite the worldsheet coordinate for different $(\Ns,\Ns')$ considered as
\begin{align}
\label{eq:15}
    (\xi,\bar\xi) \equiv
    \begin{cases}
      (w,\bar w) & \Ns=\Ns^\prime=0;\\
      (w,\theta,\bar w,\bar\theta) & \Ns=\Ns^\prime=1;\\
      (w,\theta,\bar w) & \Ns=1,\Ns^\prime=0,
    \end{cases}
\end{align}
and introduce the following notation:
\begin{align}
\label{formula:general_measure}
  \begin{aligned}
    \int d\xi d\bar\xi
    &\equiv \int dw d\bar w \int_{(\scriptscriptstyle{\Ns,\Ns^\prime})}
    %d\theta d\bar\theta
    ,\quad
    \int_{(\scriptscriptstyle{\Ns,\Ns^\prime})}
    %d\theta d\bar\theta
    \equiv
    \begin{cases}
      1 & \Ns=\Ns^\prime=0;\\
      \int d\theta d\bar\theta & \Ns=\Ns^\prime=1;\\
      \int d\theta & \Ns=1,\Ns^\prime=0.
    \end{cases}
  \end{aligned}
\end{align}
Furthermore, we understand the covariant derivatives $D,\bar D$ as
\begin{align}
\label{formula:general_covariant_derivative}
D =
  \begin{cases}
  \partial_w, & \Ns=0;\\
  \partial_\theta + \theta\partial_w, & \Ns=1;
  \end{cases}
\qquad
\bar D =
  \begin{cases}
  \partial_{\bar w}, & \Ns'=0;\\
  \partial_{\bar\theta} + \bar\theta \partial_{\bar w}, & \Ns'=1.
  \end{cases}
\end{align}
With these notations we can combine the worldsheet actions Eqs.~(\ref{formula:bosonic_action_black_hole}), (\ref{formula:typeII_action_black_hole}), and (\ref{formula:heterotic_action_black_hole}) into a single expression, read
\begin{align}
\label{formula:general_action}
  S = -\frac{1}{2\pi\alpha^\prime} \int d\xi d\bar\xi
  \left( r^2(UV) \sum_{i=1}^{d-1} \bar D X^i DX^i- {a(UV)\over 2} (\bar D U D V + \bar D V D U) \right),
\end{align}
and the vertex operators Eqs.~(\ref{formula:vertex_operator_V}), (\ref{formula:vertex_operator_U}), (\ref{formula:vertex_operator_SUSY_general}), and (\ref{formula:heterotic_vertex_operator}) into
\begin{align}
\label{formula:general_vertex_operator}
  \mathscr V_{2,4}(w,\bar w) = \int_{(\Ns,\Ns^\prime)}
  %d\theta d\bar\theta
  g_{2,4}(U) T_{2,4}(X) e^{\mp i(p_v U + k_{2,4} \cdot X)},\notag\\
  \mathscr V_{1,3}(w,\bar w) = \int_{(\Ns,\Ns^\prime)}
  %d\theta d\bar\theta
  g_{1,3}(U) T_{1,3}(X) e^{\mp i(p_u V + k_{1,3} \cdot X)}.
\end{align}
Finally, we can combine the Pomeron operators for different types of string, namely, Eqs.~(\ref{eq:14}), (\ref{formula:typeII_Pomeron_operator}), and (\ref{formula:heterotic_Pomeron_operator}), into a single expression, read
\begin{align}
\label{formula:general_Pomeron_operator}
  \int dw d\bar w \mathscr V_4(0,0) \mathscr V_2(w,\bar w)=c_{42} \delta(U) \Pi_{(\scriptscriptstyle{\Ns,\Ns^\prime})}\left(\frac{\alpha^\prime t_\mu}{4}\right) e^{ik\cdot X} \left(p_v \partial_w U p_v \partial_{\bar w} U\right)^{1+\frac{\alpha^\prime t_\mu}{4}}.
\end{align}
This shows that the Pomeron operator for scattering of strings of different types carries the same structure. Specifically, it is composed of the plane-wave factor, $e^{ik\cdot X}$, the tensor part, $\left(p_v \partial_w U p_v \partial_{\bar w} U\right)^{1+\frac{\alpha^\prime t_\mu}{4}}$, and the Pomeron propagator, $\Pi_{(\scriptscriptstyle{\Ns,\Ns^\prime})}(\frac{\alpha^\prime t_\mu}{4})$, with
\begin{align}
\label{formula:general_Pi_function}
  \Pi_{(\scriptscriptstyle{\Ns,\Ns^\prime})}(x) = 2\pi (1+x)^{\Ns+\Ns^\prime} \frac{\Gamma(-1-x)}{\Gamma(2+x)} e^{-i\pi(1+x)}.
\end{align}
We see that the string type, i.e. ($\Ns,\Ns^\prime$), enters only into $\Pi_{(\scriptscriptstyle{\Ns,\Ns^\prime})}$ (and the unimportant numerical factor $c_{42}$). Note that, notwithstanding this common structure, the poles of Eq.~(\ref{formula:general_Pi_function}) depends on ($\Ns,\Ns^\prime$), and thus the Pomeron spectrum depends on ($\Ns,\Ns^\prime$).

\section{Eikonal phase and Pomeron Regge spin}
\label{sec:eikonal_phase_string_scattering}

The general expression of the Pomeron operator, namely, Eq.~(\ref{formula:general_Pomeron_operator}), allows us to study the properties of the high-energy short-string scattering near the black hole, and to compare these properties for different string types. This is the purpose of this section. In particular, we shall calculate explicitly the eikonal phase and the Regge spin of Pomeron exchanged during string scattering.

\subsection{Eikonal phase from Pomeron operator}
\label{sec:generalities}

First of all, similar to the derivations of Eqs.~(\ref{OPE_mu_squared}), (\ref{formula:typeII_OPE_Grassmann_integrated}), and (\ref{formula:heterotic_OPE_Grassmann_integrated}), we find that for different types of strings,
\begin{align}
\label{formula:in_state_Pomeron_operator}
  \mathscr V_3(1,1) \mathscr V_1(\infty,\infty) = c_{31} e^{ip_u (\partial_w + \partial_{\bar w})V} e^{ik^\prime \cdot X} \delta(V),\qquad k^\prime = k_3-k_1,
\end{align}
where the operators $X^i,V$ on the right-hand side are evaluated at $w=\bar w=1$ and $c_{31}$ arises from $T_1$ and $T_3$. In deriving this result we have used the (super)conformal symmetry. Substituting Eqs.~(\ref{formula:general_Pomeron_operator}) and (\ref{formula:in_state_Pomeron_operator}) into the general expression of the tree-level scattering amplitude Eq.~(\ref{formula:amplitude_four_point_string}), which holds for different types of strings, we find the tree-level amplitude of string type $(\Ns,\Ns^\prime)$ in the Regge limit, read
\begin{eqnarray}
\label{eq:3}
  a^{tree}_{\scriptscriptstyle{(\Ns,\Ns^\prime)}}(s,t_\mu)=\delta^{d-1}(k_1+k_2-k_3-k_4) c_{42}c_{31}{\cal A}_{\scriptscriptstyle{(\Ns,\Ns^\prime)}}(s,t_\mu),
\end{eqnarray}
with
\begin{equation}
\label{eq:2}
  {\cal A}_{\scriptscriptstyle{(\Ns,\Ns^\prime)}} (s,t_\mu) = \frac{g^2}{\pi} \Pi_{\scriptscriptstyle{(\Ns,\Ns^\prime)}}\left(\frac{\alpha't_\mu}{4}\right) \left(\frac{\alpha's}{4}\right)^{2+\frac{\alpha't_\mu}{2}}.
\end{equation}
It is clear that $a^{tree}_{\scriptscriptstyle{(\Ns,\Ns^\prime)}}$ depends on $(\Ns,\Ns^\prime)$.

From the scattering amplitude we can follow the canonical procedures to obtain the eikonal phase:
\begin{equation}\label{eq:5}
\delta(s,b) = \frac{1}{2s} \int \frac{d^{d-1}k}{(2\pi r(0))^{d-1}} e^{ik\cdot b} {\cal A}_{(\Ns,\Ns^\prime)} (s,t_\mu).
\end{equation}
It is important that, due to $t_\mu = -(k^2+\mu^2)/r^2(0)$, this expression of eikonal phase involves a characteristic length scale, $\mu^{-1}$, which is set by the dilaton and becomes infinite only in the flat spacetime limit. So, combining $\alpha's\gg 1$, that defines the high-energy string scattering regime, with the inequality (\ref{formula:range_small_string_mu}), we find
\begin{equation}\label{eq:7}
  s^{-1}\ll\alpha'\ll \left(\frac{r(0)}{\mu}\right)^{2}.
\end{equation}

Then, for all values of $\Ns,\Ns^\prime$ considered in this work, we have
\begin{equation}\label{eq:1}
  \Pi_{(\Ns,\Ns^\prime)}(x\rightarrow 0) = -2\pi\frac{e^{-i\pi x}}{x}.
\end{equation}
This pole shows that the Pomeron spectrum includes the graviton, irrespective of the SUSY numbers. Thanks to this, by taking into account Eq.~(\ref{formula:range_small_string_mu}) we can simplify Eq.~(\ref{eq:2}) to
\begin{eqnarray}
{\cal A}_{(\Ns,\Ns^\prime)}(s,t_\mu) &\simeq& \frac{g^2}{\pi} \left(-\frac{2\pi}{\frac{\alpha^\prime t_\mu}{4}}\right) \left(\frac{\alpha^\prime s}{4}\right)^2\left(e^{-\frac{i\pi}{2}}\frac{\alpha^\prime s}{4}\right)^{\frac{\alpha^\prime t_\mu}{2}}\nonumber\\
&=&\frac{g^2\alpha^\prime s^2}{2} \frac{r^2(0)}{k^2+\mu^2} \left(e^{-\frac{i\pi}{2}}\frac{\alpha' s}{4}\right)^{-\frac{\alpha^\prime}{2r^2(0)}(k^2+\mu^2)}
\label{eq:4}
\end{eqnarray}
for $k\ll \sqrt{r^2(0)/\alpha'}$. With its substitution into Eq.~(\ref{eq:5}) we obtain
\begin{eqnarray}
\delta(s,b)\simeq \frac{g^2\alpha^\prime s}{4r^{d-3}(0)}
\int \frac{d^{d-1}k}{(2\pi)^{d-1}} \frac{e^{ik\cdot b}}{k^2+\mu^2} \left(e^{-\frac{i\pi}{2}}\frac{\alpha^\prime s}{4}\right)^{-\frac{\alpha^\prime}{2r^2(0)}(k^2+\mu^2)}.
\label{eq:6}
\end{eqnarray}
With the introduction of the following integral:
\begin{align}
I_{d-1}(a,b,\mu) \equiv \int \frac{d^{d-1}k}{(2\pi)^{d-1}} \frac{e^{ik\cdot b-a(k^2+\mu^2)}}{k^2+\mu^2},
\label{eq:8}
\end{align}
we can rewrite Eq.~(\ref{eq:6}) in a compact form, read
\begin{equation}
\delta(s,b) = \frac{4\pi G s}{r^{d-3}(0)}I_{d-1}\left(\frac{\alpha^\prime}{2r^2(0)}(\ln\frac{\alpha^\prime s}{4}-i\frac{\pi}{2}),b,\mu\right),
\label{formula:eikonal_phase_GN}
\end{equation}
where Newton's constant $G = \frac{g^2\alpha^\prime}{16\pi}$ \cite{Veneziano1987}. It is important that this expression has no dependence on $\Ns,\Ns^\prime$.

Below we discuss the behaviors of $\delta(s,b)$ in different regimes for given $\mu^{-1}$. Note that Eq.~(\ref{formula:eikonal_phase_GN}) involves a new parameter, $\alpha^\prime\ln\frac{\alpha^\prime s}{4}$, which is logarithmically larger than $\alpha'$ (because of $\alpha^\prime s\gg 1$). So, when the condition (\ref{eq:7}) is satisfied, we can still have different situations, in particular, $(\frac{r(0)}{\mu})^{2} \gg \alpha^\prime \ln \frac{\alpha^\prime s}{4}$ and of $(\frac{r(0)}{\mu})^{2} \ll \alpha^\prime \ln \frac{\alpha^\prime s}{4}$. It turns out that the behaviors of $I_{d-1}$ are very different in these two cases (cf.~Appendix \ref{sec:technical_details}). Therefore, we shall discuss the behaviors of $\delta(s,b)$ in the following two regimes separately:
\begin{itemize}
  \item High-energy scattering regime: $\alpha^\prime\ll\alpha^\prime\ln \frac{\alpha^\prime s}{4}\ll \left(\frac{r(0)}{\mu}\right)^{2}$;
  \item Ultrahigh-energy scattering regime: $\alpha^\prime\ll \left(\frac{r(0)}{\mu}\right)^{2}\ll\alpha^\prime\ln \frac{\alpha^\prime s}{4}$.
\end{itemize}
$\sqrt{\alpha^\prime\ln \frac{\alpha^\prime s}{4}}$ introduces a characteristic length scale, which characterizes the transverse
size of strings undergoing scattering \cite{BPST07}. Note that the condition: $s^{-1}\ll\alpha^\prime$ in the inequality (\ref{eq:7}) is met automatically because of $\ln \frac{\alpha^\prime s}{4}\gg 1$.

\subsection{High-energy scattering}
\label{sec:regime_I}

First of all, by definition of the high-energy scattering regime $\sqrt{s}$ cannot exceed some exponentially large energy, i.e.
\begin{align}
\left(\frac{1}{\sqrt{\alpha^\prime}}\ll\right) \sqrt{s} \ll \frac{2}{\sqrt{\alpha^\prime}} e^{\frac{c_1r^2(0)}{\alpha^\prime \mu^2}},
\label{eq:11}
\end{align}
where $c_1={\cal O}(1)$ is some irrelevant numerical constant. Below we study the behaviors of $\delta$ in distinct regimes of $b$.

For $b\gg \mu^{-1}\,(\gg\sqrt{\frac{\alpha^\prime}{r^2(0)}\ln \frac{\alpha^\prime s}{4}})$ we use Eq.~(\ref{eq:96}) to reduce Eq.~(\ref{formula:eikonal_phase_GN}) to
\begin{align}
\delta(s,b) = \frac{2\pi G s}{r^{d-3}(0)} \frac{\mu^{\frac{d-4}{2}}}{(2\pi b)^{\frac{d-2}{2}}}e^{-\mu b}\label{formula:eikonal_phase_largeb}.
\end{align}
This was found previously for bosonic (i.e. $\Ns = \Ns^\prime = 0$) string scattering, but under a different parametric condition \cite{Stanford15}. For $\mu^{-1}\gg b\, (\gg \sqrt{\frac{\alpha^\prime}{r^2(0)}\ln \frac{\alpha^\prime s}{4}})$ we can use Eq.~(\ref{eq:97}) to reduce Eq.~(\ref{formula:eikonal_phase_GN}) to (for $d>3$)
\begin{align}
\delta(s,b) = \frac{\Gamma(\frac{d-3}{2})}{\pi^{\frac{d-3}{2}}} \frac{Gs}{(r(0) b)^{d-3}}.
\label{formula:eikonal_phase_smallb}
\end{align}
This implies that, essentially, for $\mu^{-1}\gg b$ the gravitational scattering of strings recovers that of classical point particles described by the general relativity, and the eikonal phase is the Shapiro time delay \cite{Maldacena16a}. This result was not reported in the previous study for bosonic string scattering \cite{Stanford15}.

Both Eqs.~(\ref{formula:eikonal_phase_largeb}) and (\ref{formula:eikonal_phase_smallb}) describe the Regge behavior:
\begin{equation}\label{eq:16}
  \delta(s,b) \sim Gs
\end{equation}
for the gravitational scattering of classical point particles. It should be remarked that here the condition (\ref{eq:11}) is required: Only if a string shrinks to a point particle, i.e. $\alpha^\prime \rightarrow 0$, or the spacetime is flat, i.e. $\mu\rightarrow 0$, can the condition (\ref{eq:11}) and Eq.~(\ref{eq:16}) hold for arbitrarily large $s$.

\subsection{Ultrahigh-energy scattering}
\label{sec:regime_II}

By definition of the ultrahigh-energy scattering regime $\sqrt{s}$ must be extremely large so that
\begin{align}
\sqrt{s}\gg \frac{2}{\sqrt{\alpha^\prime}} e^{\frac{c_1r^2(0)}{\alpha^\prime \mu^2}}\label{formula:condition_largemu}.
\end{align}
Then, with the help of Eq.~(\ref{eq:10}) we reduce Eq.~(\ref{formula:eikonal_phase_GN}) to
\begin{align}
\delta(s,b)=\frac{4\pi G s}{r^{d-3}(0)} \left(\frac{\alpha^\prime s}{4}e^{-\frac{i\pi}{2}}\right)^{-\frac{\alpha^\prime \mu^2}{2r^2(0)}} \frac{e^{-\frac{b^2}{\frac{2\alpha^\prime}{r^2(0)}\ln \frac{\alpha^\prime s}{4}}}}{\mu^2 \left(\frac{2\pi \alpha^\prime}{r^2(0)} \ln\frac{\alpha^\prime s}{4}\right)^{\frac{d-1}{2}}}
\label{formula:eikonal_phase_largemu}
\end{align}
for $b \ll \frac{\mu \alpha^\prime}{r^2(0)}\ln \frac{\alpha^\prime s}{4}$. This result holds for different values of ($\Ns,\Ns'$) considered in this work. In the case of $\Ns=\Ns^\prime=0$ it was found in \cite{Stanford15}.

Equation (\ref{formula:eikonal_phase_largemu}) exhibits the Regge behavior, i.e.
\begin{align}
\delta(s,b) \sim Gs^{1 - \frac{\alpha^\prime \mu^2}{2r^2(0)}}.
\label{eq:24}
\end{align}
Compared to Eq.~(\ref{eq:16}), we find the Pomeron Regge spin
\begin{align}
j_0 = 2 - \frac{\alpha^\prime \mu^2}{2r^2(0)}.
\label{eq:33}
\end{align}
Thus the Pomeron Regge spin is suppressed from $2$, i.e. the classical graviton spin, by an amount of $\frac{\alpha^\prime \mu^2}{2r^2(0)}$. This stringy correction vanishes in flat spacetime. Most importantly, it has no dependence on ($\Ns,\Ns'$).
%Also, we see that the overall coefficient of $\delta$ depends on $s$ only logarithmically, and vanishes for $b \gg \sqrt{\frac{\alpha^\prime}{r^2(0)}\ln \frac{\alpha^\prime s}{4}}\,(\gg\mu^{-1})$.

\section{A theoretical model for long-string scattering in black hole background}
\label{sec:scattering_long_strings}

In this section we will go beyond the leading stringy correction (\ref{eq:33}) and study the behaviors of $j_0$ for long-string scattering in the same black hole background. In this case, because the short-string condition (\ref{formula:range_small_string_mu}) is not satisfied and instead we have
\begin{align}
\label{formula:range_long_string_mu}
\frac{\alpha^\prime \mu^2}{r^2(0)} \gg 1,
\end{align}
the framework developed in Secs.~\ref{sec:pomeron_string_scattering} and \ref{sec:eikonal_phase_string_scattering} ceases to work. Rigorous studies of high-energy scattering of two long closed strings in curved spacetime remain challenging. So below we introduce a theoretical model and give a heuristic derivation of $j_0$.

Let us make an observation on Eq.~(\ref{eq:33}). Naively, its right-hand side vanishes for long string with length $\sqrt{\alpha^\prime} = 2r(0)/\mu$. This conclusion is certainly incorrect, because for such long strings one has to take all higher order stringy corrections into account. Rather, it implies that two long strings cannot exchange a graviton during scattering, when the curvature effects of spacetime are strong. Indeed, since the Pomeron propagator $\Pi_{(\scriptscriptstyle{\Ns,\Ns^\prime})}(\frac{\alpha^\prime t_\mu}{4})$ in Eq.~(\ref{formula:general_Pomeron_operator}) exhibits a pole at $k=\pm i\mu$, the Pomeron cannot propagate in curved spacetime as freely as in a flat spacetime. Thus in the presence of strong curvature effects we expect a long string to ``split'', automatically, into $\sim \frac{\sqrt{\alpha^\prime}}{r(0)/\mu} %= \sqrt{\frac{\alpha^\prime \mu^2}{r^2(0)}}
$ uncorrelated short strings, even though the whole string does not break. (This is much like the disordered phase in the ordinary $\sigma$ model field theory, where the correlation length is finite and within that length the system is still ordered \cite{Fradkin13}.) Since it is not favored for an elastic wave --- arising from the string tension --- to travel the whole string and return back to its departure point, during scattering each short string may be regarded as an open string. In addition, such open string is either bosonic or type I. The motion of the open string is described by a worldsheet (super)conformal field theory in flat spacetime, with the same tension, i.e. $\alpha^\prime$. The corresponding metric is
\begin{equation}
\label{eq:61}
ds^2 = -a(0) dUdV + r^2(0) \sum_{i=1}^{d-1} dX^i dX^i,
\end{equation}
obtained by sending $UV$ to zero in Eq.~(\ref{formula:metric_black_hole}).

For open bosonic string the worldsheet action is
\begin{eqnarray}
\label{eq:17}
S=-\frac{1}{4\pi\alpha^\prime} \int dwd\bar w \left(r^2(0) \sum_{i=1}^{d-1} \partial_w X^i \partial_{\bar w} X^i
- {a(0)\over 2} (\partial_w U\partial_{\bar w} V + \partial_{\bar w}U\partial_w V)\right).
\end{eqnarray}
Note that $w$ is now restricted on the upper half complex plane with boundary (namely, the real axis).
%, which is denoted as $\bar{\mathbb{H}}$.
The vertex operators are placed on the boundary where $w={\bar w}$. So %Eqs.~(\ref{formula:vertex_operator_V}) and (\ref{formula:vertex_operator_U}) are modified to
%\begin{align}
%\label{formula:vertex_operator_UV_open}
%\mathscr V_{2,4}(w) &= g_{2,4}(U) T_{2,4}(X) e^{\mp i(p_v U + k_{2,4}\cdot X)},\notag\\
%\mathscr V_{1,3}(w) &= g_{1,3}(V) T_{1,3}(X) e^{\mp i(p_u V + k_{1,3}\cdot X)},
%\end{align}
%where all operators $U,X$ depend on $(w,\bar w)$. In addition,
they depend only on $w$. The operators $\mathscr V_{1,2,3,4}$ are defined in a way similar to Eqs.~(\ref{formula:vertex_operator_V}) and (\ref{formula:vertex_operator_U}), with the detailed form of $T_{1,2,3,4}(X)$ modified. The latter modifications do not affect the results below. Thus we shall not discuss them further. $\mathscr V_{1,2,3,4}$ have conformal weight $1$. Similar to the derivations in Sec.~\ref{sec:pomeron_string_scattering}, we obtain the Pomeron operator
\begin{eqnarray}
\label{eq:95}
\int dw \mathscr V_4(0) \mathscr V_2(w)\sim\delta(U) \Gamma\left(-1+{\alpha^\prime k^2\over r^2(0)}\right)e^{-i\pi(1-\alpha^\prime k^2/r^2(0))}e^{ik\cdot X} (p_v \partial_w U)^{1-\alpha^\prime k^2/r^2(0)}\quad
\end{eqnarray}
and the Reggeized scattering amplitude
\begin{equation}
\label{eq:90}
  {\cal A}_{\scriptscriptstyle{\Ns=0}} (s,-k^2)
  \sim \Gamma\left(-1+{\alpha^\prime k^2\over r^2(0)}\right)(e^{-i\pi}\alpha's)^{1-\alpha^\prime k^2/r^2(0)},
\end{equation}
where the superscript $\Ns$ denotes the SUSY number of an open string.

Similarly, for open type I superstring we have the worldsheet action
\begin{align}
\label{formula:type_I_action_black_hole}
  S = -\frac{1}{4\pi\alpha^\prime} \int dwd\bar w d\theta d\bar\theta \left(r^2(0)\sum_{i=1}^{d-1} DX^i \bar DX^i
  - {a(0)\over 2}(D U\bar D V + D V \bar D U)\right).
\end{align}
The Pomeron operator is
\begin{align}
\label{formula:type_I_Pomeron_operator}
  &\int dw {\mathscr V}_4(0) {\mathscr V_2(w)}\notag\\
  \sim &\,\,\delta(U)\left(1-{\alpha^\prime k^2\over r^2(0)}\right)\Gamma\left(-1+{\alpha^\prime k^2\over r^2(0)}\right)e^{-i\pi(1-\alpha^\prime k^2/r^2(0))}e^{ik\cdot X} (p_v \partial_w U)^{1-\alpha^\prime k^2/r^2(0)},
\end{align}
which is the same as that obtained in \cite{Cheung10} for a flat spacetime background. From this operator the Reggeized scattering amplitude
\begin{equation}
\label{eq:91}
  {\cal A}_{\scriptscriptstyle{\Ns=1}} (s,-k^2)
  \sim \left(1-{\alpha^\prime k^2\over r^2(0)}\right)\Gamma\left(-1+{\alpha^\prime k^2\over r^2(0)}\right)(e^{-i\pi}\alpha's)^{1-\alpha^\prime k^2/r^2(0)}
\end{equation}
follows.

Formally, both Pomeron propagators (\ref{eq:95}) and (\ref{formula:type_I_Pomeron_operator}) exhibit a massless pole at $k=0$, i.e. ${\cal A}_{\Ns}\sim k^{-2}s^{1-\alpha^\prime k^2/r^2(0)}$. However, such a massless Pomeron cannot exist, because the spacetime is flat only locally and the would-be massless Pomeron must attenuate at large scales. To take into account this attenuation phenomenologically, we introduce an imaginary part, $\kappa$, to the transverse momentum $k$ exchanged during the scattering. So for $k\rightarrow 0$ the amplitudes behave as
\begin{equation}
\label{eq:92}
  {\cal A}_{\scriptscriptstyle{\Ns=0,\,1}}
  {\sim} {1\over (k+i\kappa)^2}(\alpha^\prime s)^{1-\alpha^\prime (k+i\kappa)^2/r^2(0)}\sim s^{1+\alpha'\kappa^2/r^2(0)}\, {(\alpha^\prime s)^{-\alpha^\prime (k^2+2i\kappa k)/r^2(0)}\over (k+i\kappa)^2}.
\end{equation}
$\kappa$ may be identified as the mass of the Pomeron. Since the mass of the entire string may be estimated as $\sim r(0)/\sqrt{\alpha'}$ and, as discussed above, the long string splits into $\sim \frac{\sqrt{\alpha^\prime}}{r(0)/\mu}$ uncorrelated short strings, we have
\begin{equation}\label{eq:30}
  \kappa=c_2\frac{r(0)/\sqrt{\alpha'}}{{\sqrt{\alpha^\prime}/{r(0)\over\mu}}}=c_2\frac{r^2(0)}{\alpha'\mu},
\end{equation}
where $c_2$ is some numerical constant order of unity.

With the help of Eq.~(\ref{eq:92}) we can find the eikonal phase
\begin{eqnarray}
\delta(s,b)&\sim& s^{\alpha'\kappa^2/r^2(0)}\, \int \frac{d^{d-1}k}{(2\pi)^{d-1}} \frac{e^{ik\cdot b}}{(k+i\kappa)^2}\, (\alpha^\prime s)^{-\alpha^\prime (k^2+2i\kappa k)/r^2(0)}\nonumber\\
&\sim& s^{\alpha'\kappa^2/r^2(0)}\, \frac{e^{-{\left(b-\frac{2\kappa\alpha^\prime}{r^2(0)}\ln (\alpha^\prime s)\right)^2/\frac{4\alpha^\prime}{r^2(0)}\ln (\alpha^\prime s)}}}{b^2 \left(\frac{\alpha^\prime}{r^2(0)} \ln(\alpha^\prime s)\right)^{\frac{d-1}{2}}}.
\label{eq:94}
\end{eqnarray}
This yields a Pomeron Regge spin
\begin{equation}\label{eq:93}
  j_0 = 1+\alpha'\kappa^2/r^2(0).
\end{equation}
Substituting Eq.~(\ref{eq:30}) into it we obtain
\begin{equation}\label{eq:21}
  j_0 \stackrel{\frac{2r^2(0)}{\alpha' \mu^2}\ll 1}{\longrightarrow} 1+c_2^2\frac{r^2(0)}{\alpha'\mu^2}.
\end{equation}
Interestingly, in QCD $j_0\approx 1$ implies a slightly massive gauge boson, which is commonly described by the $\Ns=4$ super-Yang-Mills theory \cite{BPST07}. However, Eq.~(\ref{eq:21}) does not fall into this paradigm. The reasons are as follows: If this were true, one can read off the 't Hooft coupling in the hidden super-Yang-Mills theory from Eq.~(\ref{eq:33}), which is
\footnote{This follows directly from comparing Eq.~(\ref{eq:33}) with the large t' Hooft coupling result \cite{BPST07} $j_0=2-\frac{2}{\sqrt{\lambda}}$, where $\lambda$ is the t' Hooft coupling.}
$(\frac{4 r^2(0)}{\alpha^\prime \mu^2})^2$. This in turn gives \footnote{This follows from substituting $\lambda=(\frac{4 r^2(0)}{\alpha^\prime \mu^2})^2$ into the one-loop result $j_0=1+{\ln 2\over \pi^2}\lambda$ for small $\lambda$, obtained by using the Balitsky-Fadin-Lipatov-Kuraev (BFKL) technique \cite{Lipatov76,Kuraev77,Balitsky78}.} $j_0 = 1 + {\ln 2\over\pi^2} (\frac{4 r^2(0)}{\alpha^\prime \mu^2})^2$ for long string, but this correction is smaller than that in Eq.~(\ref{eq:21}) by one order.

\section{Pomeron Regge spin correspondence of non-maximal chaos exponent}
\label{sec:chaotic_exponent}

With the results obtained in the last two sections, we are ready to implement the general discussions in Sec.~\ref{sec:general_description}, and interpret from the perspective of string theory the scaling law of the chaos exponent of SYK-like models and its independence of SUSY(-like) structures.

First of all, observing Eqs.~(\ref{eq:33}) and (\ref{eq:21}) we find that in the short- and long-string limit, i.e. in the regimes of $\frac{r^2(0)}{\alpha' \mu^2}\gg 1$ and $\frac{r^2(0)}{\alpha' \mu^2}\ll 1$, the Pomeron Regge spin $j_0$ is determined by the same parameter, $\frac{r^2(0)}{\alpha' \mu^2}$. For the crossover regime it is difficult  to find $j_0$ analytically. However, since we cannot find other dimensionless parameters to control the behaviors of $j_0$, for arbitrary string length $j_0$ is still determined by $\frac{r^2(0)}{\alpha' \mu^2}$. So a single-parameter scaling law for $j_0$,
\begin{equation}\label{eq:99}
  j_0=1+{\tilde v}\left(\frac{4r^2(0)}{\alpha' \mu^2}\right),
\end{equation}
follows. Here ${\tilde v}(x)$ is some scaling function which, in accordance with Eqs.~(\ref{eq:33}) and (\ref{eq:21}), has the following limiting behaviors:
\begin{eqnarray}\label{eq:100}
  \tilde{v}(x)=\left\{\begin{array}{ll}
                        1-{2\over x}, & {\rm for}\, x\rightarrow\infty \\
                        {c^2_2\over 4}\,x, & {\rm for}\, x\rightarrow 0
                      \end{array}
  \right..
\end{eqnarray}
For the convenience below we introduce the factor $4$ in the definition of the scaling factor.

Next, comparing the scaling functions $v(x)$ and $\tilde{v}(x)$ for $x\rightarrow \infty$, namely, the first line of Eqs.~(\ref{eq:101}) and (\ref{eq:100}), we find that they are exactly the same. This limiting scaling behavior gives
\begin{equation}\label{eq:26}
  {\lambda_L\over {2\pi/\beta}}\stackrel{\beta\Js\gg 1}{\longrightarrow}1-\frac{2}{\beta\Js}
\end{equation}
on the SYK side and Eq.~(\ref{eq:33}) on the string theory side. The two results, when combined with the following parameter correspondence:
\begin{equation}\label{eq:27}
  \frac{\alpha' \mu^2}{4r^2(0)} \Leftrightarrow \frac{1}{\beta\Js},
\end{equation}
justify the chaos exponent-Pomeron Regge spin correspondence (\ref{eq:19}) up to the leading order $1/x$-expansion. Furthermore, because Eq.~(\ref{eq:33}) is independent of $(\Ns,\,\Ns')$, in the perturbative (strong interaction/short string) regime: $x\gg 1$ the universality of $v(x)$ with respect to SUSY structures finds its analog as the universality of $\tilde{v}(x)$ with respect to string types.

Then, let us compare the scaling functions $v(x)$ and $\tilde{v}(x)$ for $x\rightarrow 0$, i.e. in the nonperturbative (weak interaction/long string) regime. From the second line of Eqs.~(\ref{eq:101}) and (\ref{eq:100}), we find that provided $c_2$ is set to $2/\sqrt{\pi}$, they are exactly the same also. This justifies the correspondence (\ref{eq:19}) in the nonperturbative regime, and relates the universal properties of $v,\tilde{v}$ in the same fashion as in the perturbative regime.

Finally, since the behaviors of $v,\tilde{v}$ in the limiting regimes of $x\rightarrow 0$ and $x\rightarrow \infty$ are exactly the same, it is natural to expect that the $1/x$-expansion of $v,\tilde{v}$, term by term, are exactly the same also, and the universal properties of each term are related in the same fashion as before. Therefore, for arbitrary $x$ we have
\begin{equation}\label{eq:103}
  \tilde{v}(x)=v(x),
\end{equation}
which then gives Eq.~(\ref{eq:60}). This leads to, for arbitrary interaction/string length, the correspondence (\ref{eq:19}) and the interpretation of the universality of Eq.~(\ref{formula:lya_exponent}) with respect to SUSY(-like) structures in terms of the universality of Eq.~(\ref{eq:60}) with respect to string types. Under this correspondence, the flow of $\lambda_L$ from the maximal value $2\pi/\beta$ to the minimal value zero is interpreted as the flow of $j_0$ from $2$ to $1$.
%With the help of Eq.~(\ref{eq:60}), we can calculate the higher order stringy corrections to the graviton spin, which is of great interests in string theory but remains a difficult task \cite{Stanford15,BPST07}.

\section{Concluding remarks}
\label{sec:concluding_remarks}

Summarizing, we have studied the Regge spin $j_0$ of the Pomeron exchanged during high-energy string scattering in the two-sided AdS black hole. It was found that $j_0$ obeys the same single-parameter scaling law as that obeyed by the chaos exponent $\lambda_L$ of the SYK model and its variants. Strikingly, the scaling law gives a flow of $\lambda_L$ from the maximal value $2\pi/\beta$ to the minimal value zero on the SYK side, and a flow of $j_0$ from $2$ to $1$ on the string theory side. This implies that maximal chaoticity corresponds to short-string scattering and regular motion to long-string scattering. We are not aware of any reports on this correspondence. Moreover, it was found that the scaling law for $j_0$ is immune to string types, similar to the scaling law for $\lambda_L$ is immune to SUSY(-like) structures. Our findings suggest that not only maximal, but also non-maximal chaos --- even closed to completely regular motion --- in the SYK and SYK-like models may have gravity correspondence, as first noticed in \cite{Maldacena16}. Such correspondence involves strong interplay between string and black hole physics, and goes beyond the gravity correspondence of maximal chaos \cite{Maldacena15}.

Let us emphasize again that in the present work we have not solved, and have never tried to solve the challenging problem of the gravity dual of the SYK model \cite{Kitaev17,Verlinde21}. Instead, we reported a gravity analog of the non-maximally chaotic phenomenon occurring in SYK-like models reviewed in Sec.~\ref{sec:a_case_study}. Of course, it is quite possible that similar phenomena might occur to other quantum systems. In particular, by discussions in Sec.~\ref{sec:general_description} we expect the boundary system of the bulk string scattering system in the black hole background to be a good candidate. However, for string systems, even to formulate some quantum chaotic problems such as OTOC goes far beyond the present framework.

The discussions made in Sec.~\ref{sec:a_case_study} on the non-maximally chaotic phenomenon in SYK-like models crucially rely on large $q$. What happens for generic $q$? Recall that in this case a leading correction to the maximal chaos bound has been obtained for the SYK model \cite{Maldacena16}. The correction, upon rescaled by ${2\pi\over \beta}$, is $\frac{k(2)}{k_R'(-1)}\,{q\alpha_G\over \beta\Js}$ (see the original paper for the notations). However, it is very unlikely that the subleading correction would be $\propto(\frac{k(2)}{k_R'(-1)}\,{q\alpha_G\over \beta\Js})^2$, with the proportionality coefficient being independent of $q$ and $\beta \Js$. This makes us to conjecture that for generic $q$, the chaos exponent would not exhibit any single-parameter scaling behaviors. In this sense large $q$ is special. So a natural question arises: Does it have a string-theoretical interpretation? Some clues have been offered by very recent works \cite{Verlinde21,Liu23,Choi23}, but sophisticated studies are needed in order to answer this question.

The scaling law (\ref{eq:60}) for $j_0$ implies that, in the presence of curved spacetime, there is a continuous flow from $j_0=2$ to $j_0=1$. This resembles the proposal of BPST for the physics of high-energy scattering in QCD \cite{BPST07}. In that work it was proposed that both the BFKL physics for $j_0\rightarrow 1$ \cite{Lipatov76,Kuraev77,Balitsky78} and the classic Regge physics for $j_0\rightarrow 2$ \cite{Chew61,Gribov61} can be unified within the framework of a string theory in curved spacetime. This notwithstanding, the behavior for $j_0\rightarrow 1$ predicted by Eq.~(\ref{eq:60}) differs from that predicted by BFKL, as discussed in Sec.~\ref{sec:chaotic_exponent}. This might not be surprising, because the curved spacetime background in the present work differs from that in BPST's proposal. In particular, the present analysis relies on the existence of a finite-temperature black hole while BPST's analysis does not.
%{\bf check whether BPST has a black hole}

Finally, let us mention that the calculations in Secs.~\ref{sec:pomeron_string_scattering} and \ref{sec:eikonal_phase_string_scattering} can be generalized to D-branes. In view of the recent finding \cite{Verlinde21} of the relation between D-branes and the disorder-averaged SYK model, it is interesting to extend the present investigations to D-branes in the future.

\section*{Acknowledgments}

C. T. would like to thank Song He and Huajia Wang for inspired discussions on many issues touched in this work, especially string scattering in curved spacetime. He has also benefit from conversations with Eugene Bogolmony on the chaos exponent of the SYK model, and with Yibo Yang on the Regge trajectory. This work is supported by NSFC projects no. 11925507 and 12047503.

\appendix

\section{The spin-$\boldsymbol{j}$ generalization of curved spacetime scalar Laplacian}
\label{sec:spin-j_laplacian}

The OPE in this work involves vertex operators of spin-$j$ tensors. (For $j=0$ they reduce to a scalar.) Because the string propagates along either $U$ or $V$ direction, those spin-$j$ tensors have the following two special components:
\begin{equation}\label{eq:98}
  \phi_{\underbrace{\scriptstyle{U\dots U}}_{j}}(U,V,X)\equiv \phi_{\scriptstyle{U}^{j}}(U,V,X),\quad\phi_{\underbrace{\scriptstyle{V\dots V}}_{j}}(U,V,X)\equiv \phi_{\scriptstyle{V}^{j}}(U,V,X)
\end{equation}
To implement OPE in weakly curved spacetime it is necessary to introduce a covariant operator, denoted as $\Delta_j$, acting on these two fields, that generalizes the well-known scalar Laplacian
\begin{align}
\label{formula:laplacian_j=0}
\Delta_0 = \frac{1}{\sqrt{-g}} \partial_\mu \sqrt{-g} g^{\mu\nu} \partial_\nu
\end{align}
associated with the metric $ds^2=g_{\mu\nu} dx^\mu dx^\nu$, where $g$ is the determinant of the metric matrix $\{g_{\mu\nu}\}$. This issue was discussed previously for ${\rm AdS}_5$ geometry \cite{BPST07}, but for the general geometry especially with the metric  (\ref{formula:metric_black_hole}) we are not aware of any results on the explicit form of $\Delta_{j>0}$. The purpose of this Appendix is to derive this operator for the metric  (\ref{formula:metric_black_hole}).

Let us first revisit the derivations of the curved spacetime scalar Laplacian $\Delta_0$ associated with the metric (\ref{formula:metric_black_hole}). The covariant inner product of two scalar fields, $\phi(U,V,X)$ and $\phi^\prime(U,V,X)$, is defined as
\begin{align}
\label{formula:scalar_inner_product}
\int d^{d-1}XdUdV\sqrt{-g}\phi\phi^\prime,
\end{align}
where $g = g_{\parallel}g_{\perp}$ with $g_{\parallel} = -\frac{a^2}{4}$ and $g_{\perp} = r^{2(d-1)}$. Correspondingly, we have the following invariant action
\begin{align}
\label{formula:scalar_invariant_quantity}
{\cal S}_{j=0} = \int d^{d-1}dXdUdV \sqrt{-g} g^{\mu\nu}\partial_\mu \phi \partial_\nu \phi^\prime.
\end{align}
By varying this action $\Delta_{j=0}$ follows, i.e.
\begin{align}
\label{formula:scalar_invariant_into_laplacian}
&\frac{\delta}{\delta\phi^\prime} {\cal S}_{j=0} = 0\notag\\
\Rightarrow& \frac{\delta}{\delta\phi^\prime} \int d^{d-1}XdUdV\sqrt{-g}\phi^\prime \left(\frac{1}{\sqrt{-g}}\partial_\mu \sqrt{-g}g^{\mu\nu} \partial_\nu\right)\phi=0\notag\\
\Rightarrow&\Delta_0\phi = 0,
\end{align}
and with the substitution of the metric the explicit expression of $\Delta_0$ is
\begin{equation}\label{eq:13}
  \Delta_0=\frac{1}{r^2}\sum_{i=1}^{d-1}\partial_i^2-\frac{2}{ar^{d-1}}\left(\partial_Ur^{d-1}\partial_V+\partial_Vr^{d-1}\partial_U\right).
\end{equation}
Observing Eqs.~(\ref{formula:metric_black_hole}) and (\ref{formula:scalar_inner_product}), we find that $r^2$ has an effect of dilation since it is only a function of the product $UV$. This suggests the following coordinate transformation in the $U$-$V$ plane,
\begin{eqnarray}
\label{formula:UV_dilation}
&&(U,V,X)\rightarrow(U^\prime,V^\prime,X),\quad dUdV\sqrt{g_\perp} = dU^\prime dV^\prime,\nonumber\\
&&ds^2 = -\tilde a~dU^\prime dV^\prime + \tilde r^2 \sum_{i=1}^{d-1} dX^idX^i,\quad \tilde a = \frac{a}{\sqrt{g_\perp}}, \quad\tilde r = r,
\end{eqnarray}
under which both $\Delta_0$ and $\phi,\phi^\prime$ transform as a scalar.

To generalize the analysis above to $j>0$, we note that under the transformation Eq.~(\ref{formula:UV_dilation}), the spin-$j$ tensor fields considered transform as
\begin{align}
\label{formula:tensor_UV_dilation}
g_\perp^{-\frac{j}{4}}
\left\{
\begin{matrix}
\phi_{\scriptstyle{U}^{j}}\\
\phi_{\scriptstyle{V}^{j}}\\
\end{matrix}
\right\}
\rightarrow
\left\{
\begin{matrix}
{\tilde\phi}_{\scriptstyle{U}^{j}}\\
{\tilde\phi}_{\scriptstyle{V}^{j}}\\
\end{matrix}
\right\}.
\end{align}
In accordance with this transformation, we rewrite the invariant inner product
\begin{align}
\label{formula:tensor_inner_product}
\int d^{d-1}XdUdV \sqrt{-g} \phi_{\scriptstyle{U}^{j}}\phi^{\scriptstyle{V}^{j}}
\end{align}
as
\begin{align}
\label{formula:tensor_inner_product_transverse_decomposition}
\int d^{d-1}XdUdV \sqrt{-g} \left(g_\perp^{-\frac{j}{4}} \phi_{\scriptstyle{U}^{j}}\right) \left(g_\perp^{\frac{j}{4}} \phi^{\scriptstyle{V}^{j}}\right),
\end{align}
where
\begin{equation}\label{eq:104}
  \phi^{\scriptstyle{V}^{j}}(U,V,X)\equiv \phi^{\overbrace{\scriptstyle{V\dots V}}^{j}}(U,V,X).
\end{equation}
In the same way as we pass from Eq.~(\ref{formula:scalar_inner_product}) to Eq.~(\ref{formula:scalar_invariant_quantity}), we introduce the following invariant action
\begin{align}
\label{formula:tensor_invariant_quantity}
{\cal S}_{j>0} = \int d^{d-1}XdUdV \sqrt{-g} g^{\mu\nu} \partial_\mu\left(g_\perp^{-\frac{j}{4}} \phi_{\scriptstyle{U}^{j}}\right) \partial_\nu\left(g_\perp^{\frac{j}{4}} \phi^{\scriptstyle{V}^{j}}\right).
\end{align}
Then, similar to Eq.~(\ref{formula:scalar_invariant_into_laplacian}), we find the expression of $\Delta_{j>0}$ by varying ${\cal S}_{j>0}$, i.e.
\begin{align}
\label{formula:tensor_invariant_into_laplacian}
\frac{\delta}{\delta\phi^{\scriptstyle{V}^{j}}} {\cal S}_{j>0} = 0 \Rightarrow \Delta_{j>0}\phi_{\scriptstyle{U}^{j}} = 0.
\end{align}
With the substitution of Eq.~(\ref{formula:tensor_invariant_quantity}) into Eq.~(\ref{formula:tensor_invariant_into_laplacian}), we find that
\begin{align}
\label{formula:laplacian_j>0}
\Delta_{j>0} = g_\perp^{\frac{j}{4}}\,\Delta_{j=0}\,g_\perp^{-\frac{j}{4}},
\end{align}
which is the spin-$j$ generalization of the scalar Laplacian (\ref{eq:13}).

In the particular case of ${\rm AdS}_5$ geometry, Eq.~(\ref{formula:laplacian_j>0}) recovers the operator introduced in \cite{BPST07}. Let us further consider the most interesting case of $\Delta_2$ for the present geometry. Let it act on a field read $g_\perp^{\frac{1}{4}}(UV)h(X)$. With the substitution of Eq.~(\ref{formula:laplacian_j>0}), we find that near $U=V=0$ where the string scattering takes place,
\begin{align}
\label{formula:example_laplacian_j=2}
g_\perp^{-\frac{1}{4}}\Delta_2g_\perp^{\frac{1}{4}}h &= g_\perp^{\frac{1}{4}} \Delta_{0}g_\perp^{-\frac{1}{4}}h\notag\\
&= g_\perp^{\frac{1}{4}}\left(\frac{1}{r^2}\sum_{i=0}^{d-1} \partial_i^2 - \frac{4}{ar^{d-1}} \partial_U r^{d-1}\partial_V\right) g_\perp^{-\frac{1}{4}}h\notag\\
&= \frac{1}{r^2} \left(\sum_{i=0}^{d-1}\partial_i^2 + (d-1)\frac{1}{a}\partial_U\partial_V r^2\right) h\notag\\
&=-\frac{1}{r^2} \left(-\sum_{i=0}^{d-1}\partial_i^2 + \mu^2\right)h.
\end{align}
In deriving the third line we have used the fact that $r^2(UV)\approx r^2(0)+c_1UV$ for small $U,V$.
%, where the constant $c_1>0$.
The last line gives the shock wave mode of graviton first found by using the general relativity \cite{tHooft1985}.

\section{Some results of $\boldsymbol{I_{d-1}(a,b,\mu)}$}
\label{sec:technical_details}

In this Appendix, we carry out the $(d-1)$D momentum integral in $I_{d-1}(a,b,\mu)$ defined by Eq.~(\ref{eq:8}), and study its behaviors in different regimes of $a,b,$ and $\mu$. For simplicity we shall assume that $a,b,$ and $\mu$ are all real and positive.

We rewrite Eq.~(\ref{eq:8}) as
\begin{align}
I_{d-1}(a,b,\mu) = \int \frac{d^{d-1}k}{(2\pi)^{d-1}}e^{ik\cdot b - a(k^2+\mu^2)} \int_0^\infty d\lambda e^{-\lambda(k^2+\mu^2)}.
\label{eq:22}
\end{align}
Then, we exchange the order of integrals and integrate out $k$ first. As a result,
\begin{align}
I_{d-1}(a,b,\mu) = \int_0^\infty d\lambda \frac{e^{-(a+\lambda)\mu^2-\frac{b^2}{4(a+\lambda)}}}{(4\pi(a+\lambda))^{\frac{d-1}{2}}},
\label{eq:17}
\end{align}
which is rewritten as
\begin{align}
I_{d-1}(a,b,\mu) = \mu^{d-3} \int_0^\infty d\lambda \frac{e^{-(a\mu^2+\lambda)-\frac{b^2\mu^2}{4(a\mu^2+\lambda)}}}{(4\pi(a\mu^2+\lambda))^{\frac{d-1}{2}}}
\label{formula:integral}
\end{align}
with the change of variable: $\lambda\rightarrow\lambda\mu^2$.

Below we use Eq.~(\ref{formula:integral}) to derive the explicit form of $I_{d-1}$ in different regimes of $a,b,$ and $\mu$:

(i) $\sqrt{a}\ll \mu^{-1}\ll b$: We can apply the Gaussian approximation to Eq.~(\ref{formula:integral}) to obtain
\begin{align}
I_{d-1}(a,b,\mu) \simeq {\mu^{d-4\over 2}\over 2(2\pi b)^{d-2\over 2}}\,e^{-b\mu}.
\label{eq:96}
\end{align}

(ii) $\sqrt{a}\ll b\ll \mu^{-1}$: In this case, thanks to the factor: $e^{-\frac{b^2\mu^2}{4(a\mu^2+\lambda)}}$ the contribution from the regime $\lambda \lesssim a\mu^2$ is exponentially suppressed by an amount $\approx e^{-\frac{b^2}{4a}}$. The factor: $e^{-\lambda}$ further enforces the integral to be dominated by $a\mu^2\ll \lambda \lesssim 1$, for which the integrand $\approx (4\pi\lambda)^{-\frac{d-1}{2}} e^{-\lambda - \frac{b^2\mu^2}{4\lambda}}$. Taking this into account we simplify Eq.~(\ref{formula:integral}) to
\begin{align}
I_{d-1}(a,b,\mu) \simeq \mu^{d-3}\int_0^\infty d\lambda \frac{e^{-\lambda - \frac{b^2\mu^2}{4\lambda}}}{(4\pi\lambda)^{\frac{d-1}{2}}},
\end{align}
where the lower and upper integral limits have been extended to $0$ and $\infty$, respectively, since the integral in the regimes of $\lambda\lesssim a\mu^2$ and $\lambda \gtrsim 1$ are negligible. Carrying out the $\lambda$-integral \cite{Gradshteyn} explicitly gives
\begin{align}
I_{d-1}(a,b,\mu) = \frac{1}{(2\pi)^{\frac{d-1}{2}}}\left(\frac{b}{\mu}\right)^{\frac{3-d}{2}} K_{\frac{d-3}{2}}(b\mu)
\label{formula:integral_Bessel}
\end{align}
Thanks to $b\mu\ll 1$ and
\begin{align}
K_\nu(x\rightarrow 0) = 2^{\nu-1} \Gamma(\nu) x^{-\nu}, \qquad \nu\in\mathbb N /2,
\label{Bessel_identity}
\end{align}
we further simplify Eq.~(\ref{formula:integral_Bessel}) to
\begin{align}
I_{d-1}(a,b,\mu) = \frac{\Gamma\left({d-3\over 2}\right)}{4\pi^{\frac{d-1}{2}}}\frac{1}{b^{d-3}}.
\label{eq:97}
\end{align}

(iii) $\mu^{-1}\ll \sqrt{a},\,a/b$: In this case, because of $e^{-(1-b^2/(2a\mu)^2)\lambda}\approx e^{-\lambda}$ the integral is dominated by $\lambda\lesssim 1$. Thus we can simplify Eq.~(\ref{formula:integral}) to
\begin{align}
I_{d-1}(a,b,\mu) \simeq \mu^{d-3} \int_0^\infty d\lambda \frac{e^{-a\mu^2-\lambda-\frac{b^2}{4a}}}{(4\pi a\mu^2)^{\frac{d-1}{2}}}
= \frac{1}{\mu^2}\frac{e^{-\frac{b^2}{4a}-a\mu^2}}{(4\pi a)^{\frac{d-1}{2}}}.
\label{eq:10}
\end{align}

%\paragraph{Note added.} This is also a good position for notes added after the paper has been written.

\end{document}